\documentclass[12pt,preprint]{aastex}

\newcommand{\tlusty}{{\sc Tlusty}}
\newcommand{\synspec}{{\sc Synspec}}
\newcommand{\teff}{$T_{\rm eff}$}
\newcommand{\grid}{{\sc Ostar2002}}

\slugcomment{To appear in ApJS, 146 (June 2003)}

\shorttitle{NLTE Model Atmospheres of O-type Stars.}
\shortauthors{Lanz \& Hubeny}

\begin{document}

\title{A Grid of NLTE Line-Blanketed Model Atmospheres \\
       of O-type Stars}

\author{Thierry Lanz\altaffilmark{1}}
\affil{Department of Astronomy, University of Maryland, College Park, MD 20742}
\email{lanz@stars.gsfc.nasa.gov}

\and

\author{Ivan Hubeny\altaffilmark{1}}
\affil{National Optical Astronomy Observatories, Tucson, AZ 85726}
\email{hubeny@tlusty.gsfc.nasa.gov}

\altaffiltext{1}{LASP, NASA Goddard Space Flight Center, Greenbelt, MD 20771}

\begin{abstract}
We have constructed a comprehensive grid of 690 metal 
line-blanketed, NLTE, plane-parallel, hydrostatic model
atmospheres for the basic parameters appropriate to O-type stars.
The \grid\ grid considers 12 values of effective temperatures, 
$27\,500$\,K $\leq$ \teff\ $\leq\,55\,000$\,K with 2\,500\,K steps, 
8 surface gravities, $3.0\leq\log g\leq 4.75$ with  
0.25\,dex steps, and 10 chemical compositions, from metal-rich 
relative to the Sun to metal-free.  The lower limit of
$\log g$ for a given effective temperature is set by an approximate 
location of the Eddington limit.  
The selected chemical compositions have been chosen to cover a number
of typical environments of massive stars: the galactic center,
the Magellanic Clouds, Blue Compact Dwarf galaxies like I\,Zw\,18, 
and galaxies at high redshifts.  The paper contains a description
of the \grid\ grid and some illustrative examples and comparisons. The
complete \grid\ grid is available at our website 
at {\tt http://tlusty.gsfc.nasa.gov}.
\end{abstract}



\section{Introduction}
\label{Intro}

Theory of stellar atmospheres is presently enjoying a period of
revival. This renewed interest has several major reasons.

On the observational side, there has been an unprecedented expansion
of quantity and quality of stellar spectrophotometric observations.
Thanks to high sensitivity and high signal-to-noise ratio, one can
now routinely record good-quality spectra of individual stars 
in other galaxies than the Milky Way, thus giving birth to a field
of extragalactic stellar astronomy (Kudritzki 2000). The importance
of deriving accurate stellar parameters as significant  
constraints for the global galactic and cosmological parameters,
coupled with the quality of observations, compels us to use
as accurate and reliable model atmospheres as possible.

Among all stellar types, the hot, massive, and bright stars --
in particular O-type stars, stand as one of the most important
stellar groups in the global astronomical context. They contribute
most of the radiation to the UV and optical integrated spectra
of galaxies which still undergo a star-forming process. The O stars
provide the majority of ionizing photons for \ion{H}{2} regions,
and the interstellar and intergalactic medium. In the cosmological
context, there is still a debate whether hot massive stars or
Active Galactic Nuclei provided the necessary photons for the
re-ionizing of the Universe around $z \approx 3$. In view of
forthcoming space missions, like the New Generation Space Telescope,
there is a growing interest in studying the early (or even the first)
generation of stars in the Universe (see, e.g., Weiss et~al. 2000).

On the modeling side, the capabilities of the modeling methodology
has advanced enormously. One obvious reason is a many-fold
increase of computer speed and memory. Another, and also very important,
reason is the development of efficient and robust numerical
methods. The central role in this respect is played by the
Accelerated Lambda Iteration (ALI) method (for a recent review,
see Hubeny 2003).

With the development of fast computers and efficient numerical methods,
the basic limiting factor is now becoming the accuracy 
(and availability) of atomic data. Fortunately, great
advances are being made in this direction too. Recently, two major
collaborative projects,  the {\em Opacity Project} (OP) (Seaton 1987; 
The Opacity Project Team 1995; 1997), its continuation the {\em IRON Project}
(Hummer et al. 1993; Pradhan et al. 1996; Nahar 2003),
and the {\em OPAL} project \citep{OPAL91, OPAL96}, 
have produced atomic data on a large scale. 

The time is thus ripe to compute a new grid of model stellar 
atmospheres.  We need to analyze spectra of individual stars
as well as composite spectra of galaxies, with much larger degree
of complexity and accuracy  than is provided by the \citet{CD13}
spectral energy distributions (SED's) currently in use.

Because of the importance of O-type stars, such a modeling effort is 
hardly new. However, there are three issues that make a construction
of O star model atmospheres a highly non-trivial task.
They are {\sl (i)\/} the presence of a strong mass outflow, the stellar wind;
{\sl (ii)\/}  significant departures from Local Thermodynamic Equilibrium (LTE);
and {\sl (iii)\/}  an effect of numerous metal lines, traditionally
called metal line blanketing.  At present, there are no models
that would address all these three issues in full
(that is to construct fully blanketed, non-LTE models for an entire 
atmosphere, going from essentially static photosphere out to the 
wind -- the so-called unified models), although
there are several research groups and several computer programs
that are in principle capable of doing so -- programs \mbox{CMFGEN} 
(Hillier \& Miller 1998; Hillier 2003); Hoeflich's
suite of programs \citep{PH95, PH03}; PHOENIX \citep{PHOE97}; Munich 
programs \citep{Mun01}; and the Kiel-Potsdam code (Hamann 1985;
Koesterke et al. 2002).

As we will discuss in more detail in the next Section, unified models
either require huge computer resources, or one has to approximate
the other two ingredients -- NLTE and line blanketing. Also, a
description of a stellar wind is still somewhat uncertain. It is 
therefore desirable to be able to construct complete models for
hydrostatic layers, traditionally called the stellar photosphere.
Even this simplified problem is a tremendous task.
Early NLTE model atmospheres \citep{CL69, DM72} considered only
a few, about 10, energy levels of H and He. Likewise, only
most important opacity sources were considered -- several bound-free 
transition of H and He, and a few most important hydrogen lines.
On the other hand, both observations and LTE line-blanketed models 
\citep{KU79} showed that the UV spectrum contain a huge number of metal,
mostly iron, lines, which may then significantly influence not only
the emergent spectral energy distribution, but the atmospheric
structure in general. 

Although it was clear that early NLTE models provided a significantly
improved predictions of line profiles of most important spectral
lines (see, e.g., a number of papers in Garmany 1990; Crivellari, 
Hubeny, \& Hummer 1991; Heber \& Jeffery 1992)
it was not clear what is most significant from the point of view of 
an overall model atmosphere accuracy -- to compute NLTE models without
influence of metal line blanketing (or with a very simplified
blanketing), or to treat metal line blanketing as completely as
possible, while making a sacrifice of neglecting NLTE effects.

\citet{ALS85} has introduced an ingenious 
multi-frequency/multi-grey method, and was able to construct models 
with several tens to hundreds of energy levels and hundreds of lines 
treated explicitly.
When the ALI method was first implemented \citep{Wer86}, the number
of lines that can be treated explicitly increased to thousands.
\citet{ALS89} used a concept of superlevels and superlines
to describe effects of numerous lines of iron-peak elements.
Later, even more complex model atmospheres with improved
treatment of metal line blanketing were constructed
\citep{DW93, NLTE1}. The experience gained during
the last decade taught us that both NLTE and line blanketing
are very important; moreover, one needs to treat both to a very high
degree of completeness in order to avoid spurious results.

With the recent improvements of numerical schemes and with extensive 
developments of computational resources that have occurred during the 
last decade we are now able to compute essentially ``exact'',
fully blanketed NLTE hydrostatic model atmospheres.  To fulfill this 
goal is the topic of the present paper.

The paper is organized as follows. In \S\,\ref{AssMethSect} 
we summarize the basic physical assumptions and the numerical 
methods used to construct our model grid, and in \S\,\ref{BlankSect} 
we specifically describe numerical approaches to treat the metal 
line blanketing.
Section \ref{AtomSect} is devoted to an overview of atomic data, 
and, in some cases, their approximations. 
Section \ref{GridSect} describes the \grid\  model grid
in more detail. Some representative results are presented in
\S\,\ref{ResuSect}, while global parameters such as the number of
ionizing photons and a bolometric correction are discussed in
\S\,\ref{BCSect}. Section~\ref{ErrSect} discusses a sensitivity 
of computed models to adopted parameters. Our results are summarized
in \S\,\ref{ConclSect}. 


\section{Assumptions and Methods}
\label{AssMethSect}

Our basic physical assumptions are those of plane-parallel 
stratification, hydrostatic equilibrium, and radiative equilibrium.
As has been pointed out in the literature repeatedly over the past
three decades (Auer \& Mihalas 1969b; Mihalas 1978, a number of papers
in Garmany 1990; Crivellari, Hubeny, \& Hummer 1991; 
Heber \& Jeffery 1992; and Hubeny, Mihalas, \& Werner 2003), 
the approximation of Local Thermodynamic Equilibrium (LTE), 
which is another typical approximations of classical model atmospheres, 
breaks down for early-type stars, in particular for O stars.
We thus drop the approximation of LTE and compute non-LTE (NLTE) models.

\subsection{Are hydrostatic O-star model atmosphere relevant?}
\label{StaticRelev}

At the very outset, we have to clarify one crucial question.
It is well known that the most common features of the UV spectra
of O stars are strong P Cygni profiles of numerous metal lines
(\ion{C}{4}, \ion{N}{5}, \ion{O}{5}, \ion{Si}{4}, \ion{S}{4}, \ion{S}{5},
\ion{P}{5}, etc.), which
provide an unmistakable evidence of massive, supersonic outflows --
the stellar winds. Since our models are hydrostatic, we have to ask: 
are the hydrostatic model atmospheres for O stars relevant at all?

The answer depends on what exactly we are expecting from a model.
If we intend to analyze the P Cygni profiles, and to derive basic
wind parameters (mass loss rate, terminal velocity), then, obviously,
the hydrostatic models at best provide lower boundary conditions.
However, if we intend to derive the basic stellar parameters (effective
temperature, surface gravity -- therefore mass -- , and chemical composition,
from the bulk of the UV and optical spectrum), then the answer is that
the hydrostatic models are not only useful, but in many cases actually
preferable over the wind models or the unified (photosphere + wind) models
with a simplified treatment of a photosphere. There are several
reasons for this assessment.

First, most spectral lines (excluding strong, mostly resonance, lines)
are formed in the photosphere where velocities are small and geometrical
extension is negligible. Indeed, the spectrum of slowly-rotating O stars
show mostly narrow and symmetric lines (see, e.g., the high-resolution UV
spectral atlas of the O9V star, 10 Lac, Brandt et~al. 1998; and the
quality of the fit of the HST spectra by our NLTE line-blanketed 
models -- Hubeny, Heap, \& Lanz 1998). 
This statement is valid in all but the most extreme cases of
stellar winds, like those observed in Wolf-Rayet stars and extreme 
Of supergiants.
Moreover, winds are weaker in low-metallicity environments making
photospheric models even more relevant in this case.
The theoretical spectra predicted by the wind code CMFGEN \citep{CMFGEN} and by
\tlusty\  agree very well for a test case, \teff=35\,000\,K, $\log g=4.0$,
demonstrating that most lines are indeed formed in the quasi-static 
photosphere \citep{LexJDH, JCB346}.

Second, there are many remaining uncertainties as to the exact properties of
stellar winds of hot stars, even though the paradigm of radiatively-driven
winds is well established. For example, we do not have in many cases a consistent
solution: the calculated line force is often too small to drive a wind through
the critical point (though some progress is being made),
so a $\beta$-type velocity law must be adopted
to describe empirically the velocity and density structure of the wind. Furthermore,
radiative equilibrium is not really satisfied in these winds, where shocks dump
mechanical energy and heat the wind, thus changing its ionization structure
(e.g., super-ionization). Finally, the assumption of one-dimensional geometry is
challenged by the likely presence of dense clumps in the wind and by rotation;
the role of magnetic fields is  not yet understood, and their importance in structuring
the wind will very likely be more important than in the photosphere.

Third, there is also a practical aspect. Spherical model atmospheres
with velocity fields, although technically still a 1-D problem, 
have effectively a larger dimensionality than a plane-parallel static
problem because of a sharp angular dependence of the radiation field. 
(Recall that in spherical wind models, the number of angles is about 
the same as the number of discretized radial points, while in the
plane-parallel case considering a few -- typically 3 -- angles is usually 
sufficient). Consequently, computing spherical expanding models
is intrinsically more complex and more computationally intensive than
plane-parallel models. One is thus forced either to compute fewer
models, or to make approximations in the local physics (in particular
opacities), to reduce computer requirements to reasonable limits.

Finally, it is important to realize that while line blanketing
operates everywhere in the atmosphere, departures from
hydrostatic equilibrium occur only in the outer layers.  History has
taught us that it is very dangerous to neglect or even to approximate
metal line blanketing.  Neglecting line blanketing leads not only to
a lower local temperature in the continuum-forming layers and thus to
systematic errors in determining the effective temperature (e.g. Hubeny,
Heap \& Lanz 1998), but also to errors in predicted ionization balance
of important elements (CNO, etc.). These species provide in turn important
diagnostic tools for the wind. Since the ionization edges of CNO
and other light metals are located in the far and extreme UV, and
since this region is crowded with numerous lines of highly ionized
iron and nickel (\ion{Fe}{4}-\ion{Fe}{6}; \ion{Ni}{4}-\ion{Ni}{6}),
any deficiencies in describing photospheric line blanketing are
directly reflected in the uncertainties in the predicted wind structure.
It is thus no exaggeration to say that without a proper and detailed treatment
of photospheric metal line blanketing even sophisticated wind models
may be seriously in error.

Therefore, we are on much safer ground to determine the fundamental parameters
of massive stars as long as we use selected photospheric lines and avoid
all uncertainties remaining with wind models. Moreover,
a grid of NLTE line-blanketed model atmospheres incorporating state-of-the-art
opacities will provide a sound basis for comparison to future, more realistic
model atmospheres of O stars.

\subsection{Numerical Methods}
\label{NumSect}

Over the years, we have developed  a computer program called \tlusty\ 
for computing model stellar atmospheres  without assuming LTE
\citep{Tlu88, NLTE92, NLTE1, Tueb1, Lex01, IAU210}.
The program solves the basic equations -- radiative transfer,
hydrostatic equilibrium, radiative equilibrium, statistical equilibrium,
charge and particle conservation.
 
These equations are discretized in frequency and depth, which yields 
a set of highly-coupled, non-linear algebraic equations. 
We have used in this work the so-called hybrid CL/ALI method that 
was introduced by \citet{NLTE1}.
This method builds on the advantages of both ALI and Complete Linearization. Through ALI,
the mean intensities of the radiation field are eliminated from the set of linearized
equations. However, the hybrid CL/ALI method still allows us to linearize the radiation
intensity at a few selected key frequencies, notably improving the convergence
rate compared to a ``pure'' ALI approach. The hybrid CL/ALI method combines therefore
the low cost of ALI while retaining a convergence rate that is similar to CL.
Moreover, levels and superlevels may be grouped during the linearization step, 
reducing even further the number of linearized equations \citep{Tueb1}. The full
statistical equilibrium is solved in a separate step, element by element, with a
typical maximum number of levels around 200 per element.

Under the above assumptions, the critical ingredient of the models is 
a degree of completeness of the opacity sources, and of the 
atomic/ionic energy levels treated explicitly in the statistical
equilibrium equations. We shall discuss these issues in more detail in
the next two Sections.


\section{Treatment of Metal Line Blanketing}
\label{BlankSect}

As we have pointed out above, 
NLTE line blanketing is an essential ingredient of realistic model atmospheres,
determining the vertical temperature structure and the ionization structure.
NLTE ionization rates are typically larger than their LTE counterparts, because the
radiative rates are determined by the radiation field which is emitted in deeper, hotter
layers. Therefore, accurate mean intensities of the radiation field are
required to evaluate correctly the NLTE ionization structure, 
which in turn require an accounting of all sources of opacity. When
neglecting in particular the opacity of Fe lines, NLTE models predict too strong
EUV/UV radiation fields, and overestimate the ionization and the importance of
departures from LTE. In order to construct realistic NLTE model atmospheres, we have
therefore included about 100,000 individual atomic levels of over 40 ions of
H, He, C, N, O, Ne, Si, P, S, Fe, and Ni in the calculations. The levels are grouped
into about 900 NLTE superlevels, with populations of individual energy levels following
Boltzmann statistics inside a superlevel. All individual levels in a superlevel thus
share the same NLTE departure coefficient. \citet{NLTE1} discussed the concept of
NLTE superlevels in detail. A total of 8000 lines of the light elements and about
2 millions lines of Fe\,{\sc iii-vi} and Ni\,{\sc iii-vi} are accounted for in the
calculations. Based on a line strength criterion, Fe and Ni lines are dynamically selected
from a total of 12 million lines listed by \citet{CD22}. The numbers of NLTE levels
and lines included in the model atmospheres are listed in Table~\ref{IonTbl}.
Details of the model atoms are given in Sect.~\ref{AtomSect}.

ALI and level grouping are two major steps toward the practical realization
of NLTE line-blanketed model atmospheres. ``Superlines'', i.e. transitions
between superlevels, are another important ingredient. A superline may involve
hundreds or thousands of individual lines over a relatively broad frequency range.
The resulting absorption cross-section is then very complex and a detailed
representation may require a very large number of frequency points.
In an early implementation of NLTE line blanketing, \citet{NLTE1} adopted
the concept of Opacity Distribution Functions (ODFs) to describe the absorption
profile of superlines. The total opacity of all lines in a given transition
(computed using Voigt profiles for each line) is sorted; the sorted opacities could
then be represented with a limited number of frequencies, typically 15 to 30 per transition,
or 30\,000 to 50\,000 frequencies for the whole spectrum. The ODF approach assumes
that the background opacity changes little over the frequency range covered by the
superline. Details of blends with other lines or continua are therefore lost; however,
this issue should not be serious in a statistical sense since this method
is designed for millions of lines of the iron-peak elements. One may still 
unintentionally create or neglect blends between strong lines and ODFs, thus
affecting the resulting model atmospheres.

We have therefore implemented a second method, Opacity Sampling (OS), that have
the advantage of treating exactly the blends. OS is a simple Monte Carlo-like sampling
of the superline cross-sections. Contrary to the ODF approach, OS may miss
the cores of important strong lines if too few frequency points are used.
We use a variant of OS, sampling the whole spectrum at prescribed
intervals in frequency. The typical sampling step is 0.75 fiducial Doppler
widths of Fe. Smaller steps are occasionally used to include light elements' line
cores in the frequency set as well as points around photoionization edges. Our model
atmospheres are sampled with 180\,000 to 200\,000 frequency points for an O star
model. While costly in terms of computing requirements, we are thus sampling the
spectrum with sufficient resolution to ensure that every line is accounted
for. This is the current standard mode in \tlusty, which was used to compute
this grid of model atmospheres.

\citet{Tueb2} presented some illustrations of the ODF and OS approaches.
In Sect.~\ref{ErrSect}, we show the changes in a typical model atmosphere
that result from different representations of line opacities.


\section{Atomic Data}
\label{AtomSect}

The Opacity Project \citep{IOP95, IOP97} provides very extensive datasets 
for radiative transitions ($gf$-values, photoionization cross-sections) which
have been obtained from ab-initio calculations for all ions of the most abundant
light species ($Z\leq 14$, S, Ca, and Fe). The current status of the project
and its continuation as the Iron Project is discussed by \citet{NS03}.
OP claims that their data are accurate at the 10\% level or better. We have
adopted systematically these data for light elements and they form the bulk
of radiative data used in \tlusty. The OP data have been extracted from
Topbase\footnote{{\tt http://heasarc.gsfc.nasa.gov/topbase/}},
the database of the Opacity Project \citep{topbase}. OP also provides
theoretical level energies. Despite the sophistication of OP calculations,
these energies still differ by a few percent from the energies measured in the
laboratory. We have therefore updated the level energies with experimental energies
when they are available in the Atomic and Spectroscopic
Database\footnote{{\tt http://physics.nist.gov/cgi-bin/AtData/main\_asd}} at NIST \citep{NIST}.
These large datasets are manipulated with {\sc Modion}, an IDL-based graphic tool
that works directly with Topbase data files \citep{Modion}, to prepare the model
atoms required by \tlusty. For most lines, we have adopted depth-independent Doppler
line profiles, with a standard temperature, T=0.75\,\teff.
Depth-dependent, Stark or Voigt profiles are used for the strongest lines
(about 70 lines of light elements and the lower line series of \ion{H}{1} and
\ion{He}{2}). Radiative damping constants are extracted from the NIST database,
and Stark widths are adopted from the work of Dimitrijevic and collaborators.
OP photoionization cross-sections contain
many autoionization resonances. As \tlusty\  handles large frequency sets,
we could therefore adopt the theoretical OP cross-sections in every detail.
However, sharp resonances are not fully resolved by the OP calculations. Moreover,
resonances are shifted off their exact wavelengths due to the limited accuracy
of theoretical energies. Therefore, we have followed \citet{Bau98} and
implemented the concept of Resonance-Averaged Photoionization (RAP) cross-sections.
The cross-sections are smoothed by a Gaussian with a width, $\delta E/E=0.03$,
that corresponds to the typical uncertainty of the resonance energies.
OP photoionization cross-sections are total cross-sections for each initial
state summed over all final states. We have kept this approximation: we assume
that all bound-free transitions considered by \tlusty\  end up in to the ground
state of the next ion. If the transition to the ground state is forbidden by selection
rules, the cross-sections is shifted in energy corresponding to the excitation of
the lowest excited state for which the transition is allowed. 
While the opacity and the ionization rate are thus computed
correctly, the corresponding recombination rate (based on detailed balance
arguments) is only approximate because it assumes that the end state 
of ionization is in Boltzmann equilibrium with the ground state.
Nevertheless, since the population balance of the upper ion is determined
by a large number of processes (mostly bound-bound transitions), the 
influence of an approximate recombination rate is typically small,
and the resulting error in the statistical equilibrium virtually negligible.
Additional details on the model atoms for each ion is given in the relevant
subsections.

For iron-peak elements, we have extracted level energies and line oscillator
strengths from \citet{CD22} extensive semi-empirical calculations. \citet{CD22}
lists thousands of energy levels for each ion. A statistical approach to deal with
so many levels is therefore necessary; solving the full statistical equilibrium
equations would indeed require inverting very large matrices which is not only costly
but numerically inaccurate. We therefore group individual
levels into superlevels, assuming that the population of each level in a given
superlevel follow the Boltzmann distribution. While the populations of superlevels
could depart from their LTE values, all sublevels in a given superlevel share
the same $b$-factor. We have applied two criteria to group the levels: all sublevels
must have very close excitation energies, and they share the same parity. The first criterion
ensures that collisional rates between sublevels are large enough to achieve
the Boltzmann statistics; the second criterion further supports the validity of this
assumption by treating properly transitions out of and in to metastable states.
The second requirement provides the additional advantage of avoiding radiative
transitions within a single superlevel. The absorption cross-section of a superline
between two superlevels can be quite complex, involving hundreds or thousands of
individual lines. Exact absorption cross-sections are computed at three depths
(top and bottom of the model atmosphere, and at a depth where $\tau_{\rm Ross}\approx 1$
and $T\approx$\,\teff), and are logarithmically interpolated between these points.
A detailed representation requires a very large number of
frequency points. Opacity Sampling is a method designed for handling such cases,
and its implementation in \tlusty\  was outlined in Sect.~\ref{BlankSect}.

Non-LTE calculations also require collisional excitation and ionization rates.
However, collisional rates are known only for a limited set of transitions,
mostly inter-system transitions that are important for diagnostics at low densities.
The Iron Project \citep{IP93} is starting to remedy this situation, but their data
are not yet implemented in \tlusty. We rely on general formulae at this stage.
We use the \citet{Colbf62} formula for collisional ionization; for optically-allowed
line transitions, we use the \citet{Colbb62} formula which is accurate to
about 30~-~50~\% (Pradhan, priv. comm.); for forbidden transitions, no approximate
formula or a simple prescription is available, so we somewhat arbitrarily use 
Eissner-Seaton's formula with $\gamma(T)=0.05$. Different approaches of collisional
strengths in superlines may be implemented. \citet{Tueb2} compared two different
estimates and they argued that the most natural way consists in summing up all
contributions for individual transitions, using the appropriate formulae for
allowed and for forbidden transitions. We have adopted this approach here.

The following subsections provide the relevant details of the model atoms.
Table~\ref{IonTbl} summarizes the atomic data included in the model atmospheres,
as well as the references to the original calculations collected in Topbase.
Datafiles and Grotrian diagrams may be retrieved from \tlusty's Web site,
{\tt http://tlusty.gsfc.nasa.gov}.

\subsection{Hydrogen}
\label{hydrogen}

The model atom of hydrogen includes the lowest 8 levels and one superlevel
grouping the higher excitation levels, from $n=9$ to $n=80$, following the
occupation probability formalism developed by \citet{HM88} and extended
to NLTE situations by \citet{HHL94}. The Lyman and Balmer continuum
cross-sections are generalized to account for transitions from these two
levels to dissolved higher states. This contribution is neglected for all
other continua.

All lines between the first 8 bound levels are included, assuming an
approximate Stark profile for the Lyman and the Balmer series \citep{HHL94}
and a Doppler profile for the other lines. The higher members of the Lyman
and the Balmer line series are incorporated via the OS approach.
This is the only difference from the treatment of \citet{HHL94}
who adopted ODFs. OS is used here in order to account exactly
for line blending with the \ion{He}{2} line series (and with other lines).
Higher members of other line series are neglected.

Hydrogen is the main opacity source in the photospheres of O stars. We can
benefit most of our hybrid CL/ALI method by selecting a few key frequencies
where the hydrogen opacity is largest, at which the mean intensity of the
radiation field is linearized. We have selected the first six frequencies
immediately higher than the
Lyman edge, the three core frequencies in Ly~$\alpha$, and the core
frequency of Ly~$\beta$, Ly~$\gamma$, H~$\alpha$, and H~$\beta$.
A few more frequencies have been selected in the cores of strong resonance
lines of other species, and will be mentioned in the relevant subsections. 
Although there might be a different optimal choice of these frequencies for
each individual model atmosphere, we have found that this choice was generally
adequate to achieve most of the gains of the CL/ALI method in the parameter space
covered by the \grid\  grid. 

\subsection{Helium}
\label{helium}

Neutral helium is modelled with a 24-level atom, which incorporates all 19
individual levels up to $n=4$, and 5 superlevels grouping levels with
identical principal quantum numbers ($5\leq n\leq 8$); for $n=5$, singlet and
triplet levels are grouped into two separate superlevels. We have performed some
tests with more detailed model atoms, splitting the superlevels up to $n=6$,
or including higher excitation levels, but we did not find significant
differences. All lines from $n\leq 4$ levels are included, but infrared
lines between high-excitation levels ($n\geq 5$) are assumed to be in
detailed radiative balance. Voigt line profiles are used for lower-excitation
lines ($n\leq 2$), and Doppler profiles are assumed for higher-excitation
lines. \ion{He}{1} is treated in a special way in \tlusty;
like hydrogenic ions, and contrary to all other ions, we have special
subroutines in the code to deal with atomic data. We use Hummer's (priv.
comm.) routine to evaluate the photoionization cross-sections; the routine
is based on Seaton~\& Fernley's cubic fits of OP data, and properly manages
the different ways of averaging levels. Bound-free and bound-bound collisional
transitions rates are also evaluated with a special routine based on approximate
expressions from \citet{MHA75}. For bound-bound transitions between low-excitation
levels, these approximate rates are superseded by detailed calculations of
collisional cross-sections \citep{BK87}, used by Storey and Hummer (priv. comm.)
to evaluate collisional rate coefficients.

The ionized helium model atom includes the first 20 bound levels. The \ion{He}{2}
model is built following closely \ion{H}{1}. We will only indicate here the
differences. The occupation probability formalism \citep{HHL94} is taking into
account for the 20 bound levels, but higher-excitation levels are not included
explicitly. We also use continuum cross-sections for bound-free transitions from the
ground and first excited levels that account for transitions to highly-excited,
dissolved levels. Stark profiles are adopted
for the first four line series, while infrared lines ($n>10$) are set to detailed
radiative balance. Finally, the first three frequencies at the head of the \ion{He}{2}
Lyman continuum are linearized in the ALI/CL scheme.

\subsection{Carbon}
\label{carbon}

The general handling of atomic data has been discussed at the top of Sect.~\ref{AtomSect}.
We restrict the following subsections to a few particulars relative to each ion. 
Carbon is represented by a 22-level \ion{C}{2} model atom, a 23-level \ion{C}{3}
model atom, and a 25-level \ion{C}{4} model atom, grouping 44, 55, and 55 individual
levels, respectively. The \ion{C}{2} model atom includes the first 17 levels, 14 from
the doublet system and the first three levels of the quartet system. Five superlevels
group higher-excitation levels, four in the doublet system ($6\leq n\leq 9$) and one
superlevel grouping three levels in the quartet system. The \ion{C}{3} model atom
includes the first 16 levels ($E\leq 312\,000$\,cm$^{-1}$), and group all higher-excitation
levels up to $n=6$ into 7 superlevels. Levels with $n\geq 7$ are not included
explicitly (see further discussion in Sect.~\ref{ErrSect} about this restriction and
its consequence on C$^{3+}$ recombination). The \ion{C}{4} model atom includes all
individual levels up to $n=6$ and four superlevels ($7\leq n\leq 10$).
The data extracted from Topbase have been extended to include all levels up to
$10m\,^2$M$^0$. We have included the two fine-structure levels of $2p\,^2$P$^0$ for
an exact treatment of the \ion{C}{4}\,$\lambda$1549 resonance doublet.
Depth-dependent Voigt profiles are adopted for the resonance lines of the three ions,
and for the strong \ion{C}{3}\,$\lambda$1176 line from the $^3$P$^0$ metastable level.

\subsection{Nitrogen}
\label{nitrogen}

The model atmospheres incorporate four ions of nitrogen, \ion{N}{2} to \ion{N}{5},
and the ground state of \ion{N}{6}. The first 16 levels in the singlet and triplet
systems of \ion{N}{2} ($n=2$; $3s$, and $3p$ levels), and the first two quintet levels,
are included in the models as explicit individual NLTE levels. Higher-excitation
levels ($3d$; $4\leq n\leq 5$) are grouped into 8 superlevels, four in each system.
Higher levels ($n\geq 6$) are not incorporated explicitly in the model atom.
The \ion{N}{3} model atom includes all the levels below the ionization limit
as extracted from Topbase. The first 24 levels ($E\leq 330\,000$\,cm$^{-1}$) are treated
as individual levels, and the other 43 levels are grouped into 7 superlevels.
The ground state is split into two fine-structure levels for an exact treatment
of the resonance doublets (e.g., \ion{N}{3}\,$\lambda$990). The \ion{N}{4} model atom
was constructed similarly to the \ion{C}{3} model atom. The first 15 levels are included
explicitly, and higher levels up to $n=6$ are grouped into 8 superlevels. Levels
with $n\geq 7$ are not included explicitly. Finally, the \ion{N}{5} model atom is
similar to \ion{C}{4}, although levels are grouped into 6 superlevels starting with
$n=5$ levels. The two fine-structure levels of $2p\,^2$P$^0$ are included in the model atom.
Depth-dependent Voigt profiles have been set up for 18 resonance lines of the four ions.

\subsection{Oxygen}
\label{oxygen}

The model atmospheres incorporate five ions of oxygen, \ion{O}{2} to \ion{O}{6},
and the ground state of \ion{O}{7}. \ion{O}{2} and \ion{O}{3} have a relatively rich
structure of levels. High excitation levels have thus not been incorporated in the model
atoms. The \ion{O}{2} model atom includes 21 individual levels and 8 superlevels, grouping
levels up to $n=5$, while the \ion{O}{3} model atom includes the lowest 20 levels
and groups higher-excitation levels up to $n=6$ into 9 superlevels. All \ion{O}{4} below
the ionization limit are incorporated, the first 30 as individual levels, the others grouped
into 8 superlevels. Fine-structure of the \ion{O}{4} ground state is included. The
first 34 \ion{O}{5} levels are included explicitly, and higher levels up to $n=7$
are grouped into 6 superlevels.
Finally, the \ion{O}{6} model atom is similar to \ion{C}{4}, although levels are grouped into
5 superlevels starting with $n=6$ levels. The two fine-structure levels of $2p\,^2$P$^0$ are
included in the model atom. Depth-dependent Voigt profiles have been set up for
\ion{O}{4}\,$\lambda$789, 609, 554, and \ion{O}{5}\,$\lambda$630.

\subsection{Neon}
\label{neon}

Simple model atoms have been built for the neon ions, \ion{Ne}{2} to \ion{Ne}{4}.
We are not aiming at a detailed NLTE calculation of Ne populations, but our
purpose was to account for the bound-free opacity in the extreme-ultraviolet due
to the ionization of this abundant species. The \ion{Ne}{2} model atom includes
the first 11 levels and groups levels up to $E\leq 300\,000$\,cm$^{-1}$ into
four superlevels. \ion{Ne}{3} levels up to $E\leq 400\,000$\,cm$^{-1}$ are
included in the model atom, as 12 individual levels and 2 superlevels. Finally,
the \ion{Ne}{4} model atom includes the first 10 levels, and 2 superlevels
grouping levels up to $E\leq 550\,000$\,cm$^{-1}$. All lines are assumed to have
depth-independent Doppler profiles.

\subsection{Silicon}
\label{silicon}

Detailed model atoms have been constructed for \ion{Si}{3} and \ion{Si}{4}, because the
silicon ionization balance is a good temperature indicator for late O and B stars.
The \ion{Si}{3} includes all 105 energy levels listed by Topbase below the
ionization limit. The first 24 levels are treated as individual explicit levels, while
the higher levels are grouped into 6 superlevels (3 superlevels in the singlet and
in the triplet systems). The \ion{Si}{4} model atom includes all
individual levels up to $n=6$ and four superlevels ($7\leq n\leq 10$).
The data extracted from Topbase have been extended to include all levels up to
$10m\,^2$M$^0$. We have included the two fine-structure levels of $2p\,^2$P$^0$ for
an exact treatment of the \ion{Si}{4}\,$\lambda$1397 resonance doublet.
Depth-dependent Voigt profiles are adopted for \ion{Si}{3}\,$\lambda$1206, 567,
and \ion{Si}{4}\,$\lambda$1394, 1403 resonance lines.

\subsection{Phosphorus}
\label{phosphorus}

Phosphorus has not been considered by OP, because the relatively low phosphorus
abundance in the Sun indicates that phosphorus is most likely only a minor contributor
to the total opacity in astrophysical plasmas. However, we have included phosphorus
as an explicit NLTE species in our model atmospheres due to the potential diagnostic
value of the \ion{P}{4}\,$\lambda$951 and \ion{P}{5}\,$\lambda$1118-28 resonance lines.
We have built simple model atoms based on data extracted from the NIST Atomic and Spectroscopic
Database. The \ion{P}{4} model atom includes the first 14 levels, while the \ion{P}{5}
model atom includes individual levels up to $n=5$ and groups $6\leq n\leq 9$ levels
into 4 superlevels.
We have approximated the photoionization cross-sections by interpolating
in cross-sections from the same configurations
in isoelectronic sequences (e.g., using \ion{Si}{4} and \ion{S}{6} cross-sections
to estimate \ion{P}{5} cross-sections). We believe that this approach is likely more
realistic than a simple hydrogenic approximation.

\subsection{Sulfur}
\label{sulfur}

The model atmospheres incorporate four ions of sulfur, \ion{S}{3} to \ion{S}{6},
and the ground state of \ion{S}{7}. The model atoms are kept relatively simple,
aiming at including the sulfur bound-free opacity in the model atmospheres and
at predicting a few key diagnostics lines, mostly \ion{S}{4}\,$\lambda$1073 and
\ion{S}{5}\,$\lambda$1502. The \ion{S}{3} model atom includes all levels with
$E\leq 200\,000$\,cm$^{-1}$, the first 16 levels individually and the other
11 levels grouped into 4 superlevels. The \ion{S}{4} model atom includes the
first 14 levels up to $E\leq 225\,000$\,cm$^{-1}$, while the \ion{S}{5}
model atom includes the first 12 levels. The \ion{S}{6} model atom includes all
individual levels up to $n=5$ and three superlevels ($6\leq n\leq 8$).
Fine structure of \ion{S}{4} and \ion{S}{6} $3p\,^2$P$^0$ levels is taken into
account. The \ion{S}{6}\,$\lambda$933-45 resonance lines are the only sulfur
lines for which we have adopted depth-dependent Voigt profiles.

\subsection{Iron}
\label{iron}

The model atoms of iron-peak elements are based on \citet{CD22} extensive
semi-empirical calculations. We have extracted a list of even and of odd-parity
levels for each ion included in the model atmospheres. We have considered all
levels, those with experimental energies as well as those with predicted energies.
We plotted histograms of numbers of levels per energy interval to set up the energy
bands defining the superlevels. We chose typically 15 to 30 bands per parity,
so the model atoms have 30 to 50 superlevels. For each ion, we read the full
linelist from \citet{CD22} and select dynamically the strongest lines. The
line selection criteria include line $gf$-values, excitation energies,
and ionization fractions. In this way, we skip most or all \ion{Fe}{6} lines
in the coolest models and most or all \ion{Fe}{3} lines in the hottest models.
We have extracted photoionization cross-sections from the latest calculations
of \ion{Fe}{3} to \ion{Fe}{5} by the Ohio State group (see Table~\ref{IonTbl}).
They assumed $LS$-coupling, and we could typically assign theoretical
cross-sections to observed levels for the lowest 20 to 30 levels. We have assumed
an hydrogenic approximation for higher-excitation levels and for all \ion{Fe}{6}
levels. The data were then summed up to setup cross-sections for the superlevels.
Finally, we have applied to these cross-sections the RAP smoothing technique.

\subsection{Nickel}
\label{nickel}

We have followed the same approach for nickel, using \citet{CD22} data. OP did
not include nickel in their calculations. We have therefore adopted an hydrogenic
approximation for the photoionization cross-sections.


\section{Description of the Grid}
\label{GridSect}

The \grid\  grid covers the parameter space of O stars in a dense and comprehensive way.
We have selected 12 effective temperatures, $27\,500$\,K$\leq$\teff $\leq 55\,000$\,K,
with 2\,500\,K steps, 8 surface gravities, $3.0\leq\log g\leq 4.75$, with
0.25\,dex steps, and 10 chemical compositions, from metal-rich relative to the
Sun to metal-free. Table~\ref{KeyTbl} lists the selected metallicities. Solar
abundances refer to \citet{Sun98}. In every case, we have assumed a solar
helium abundance, He/H=0.1 by number. All other chemical abundances are scaled
from the solar values.  \tlusty\  can readily accommodate non solar-scaled abundances
resulting, e.g., from mixing of CNO-cycle processed material; however, this would
increase the total number of models to compute to an unmanageable level by increasing
the dimensionality of the parameter space. Thus, we view this grid as a starting point
from which detailed abundance analyses can be performed.
The selected chemical compositions have been chosen to cover a number
of typical environments of massive stars: the galactic center (``C models''),
the Magellanic Clouds (``L models'' for LMC; ``S models for SMC''),
Blue Compact Dwarf galaxies (e.g. ``W models'' for I\,Zw\,18), and galaxies
at high redshifts (``X and Y models''). 

\begin{figure}
\plotone{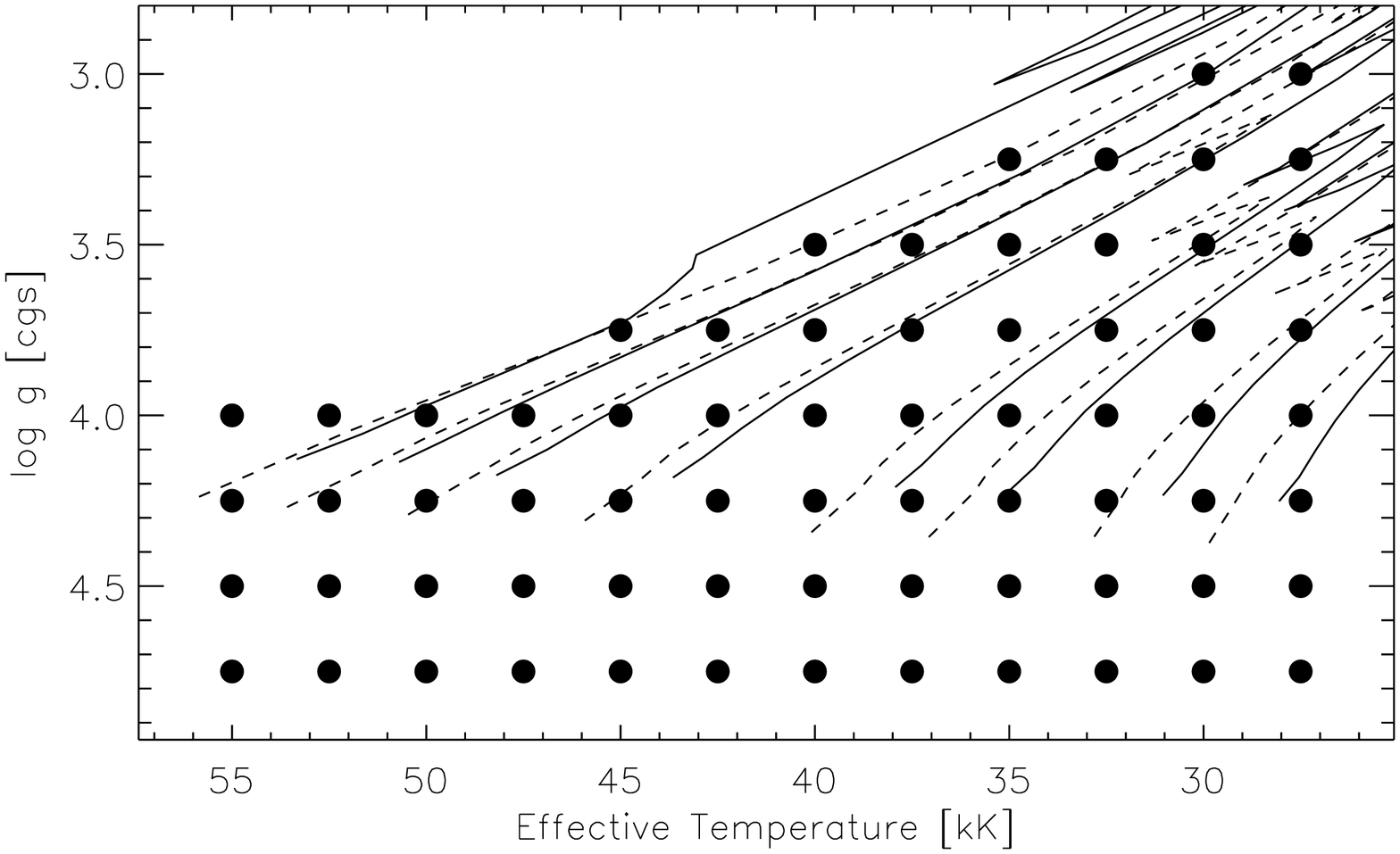}
\caption{Selected \grid\  grid points in the $\log g\  vs.$ \teff\  plane.
Geneva evolutionary tracks \citep{Geneva1, Geneva3} are shown for solar and
1/5 solar metallicities (full and dashed lines, respectively).
The tracks correspond to models with initial masses
of 120, 85, 60, 40, 25, 20, 15, 12\,$M_\sun$ from left to right, respectively. \label{TeflogFig}}
\end{figure}

For each composition, we have computed 69 model atmospheres, as illustrated
in Fig.~\ref{TeflogFig}. 
Evolutionary tracks from the Geneva group \citep{Geneva1, Geneva3} have been
overplotted in Fig.~\ref{TeflogFig}. Stars with masses larger than about
15\,$M_\sun$ are in the range covered by our model atmospheres.
The highest gravity and the lowest
temperature have been included in the grid to avoid extrapolations when using
this grid as a starting point for NLTE spectral analyses.

\begin{figure}
\plotone{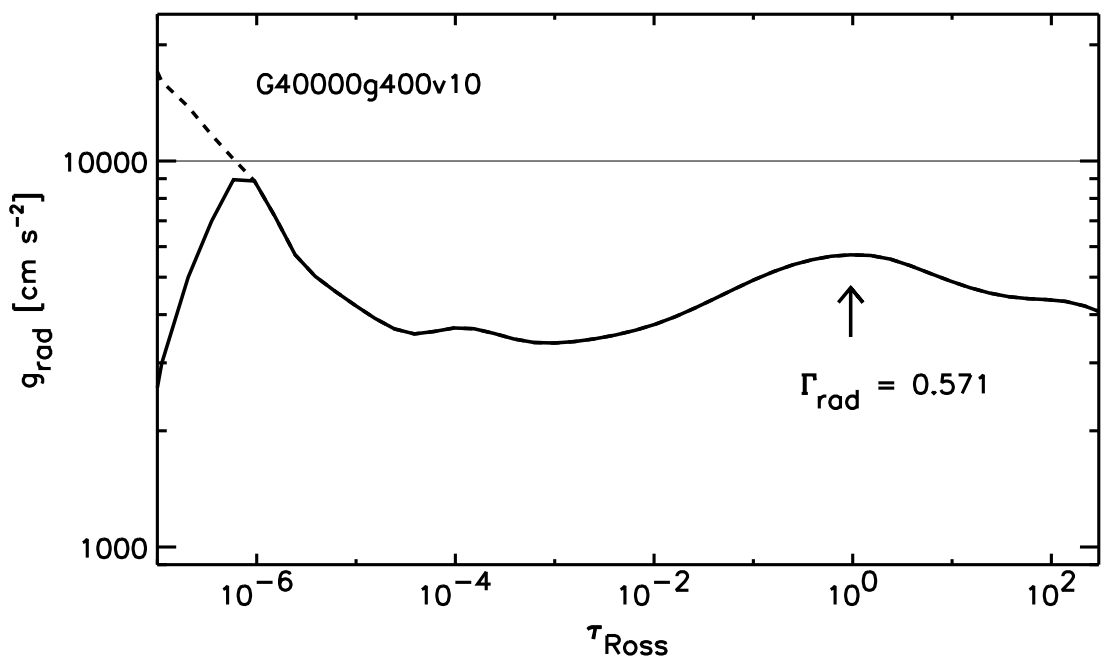}
\figurenum{2}
\caption{ Radiative acceleration (thick line) for a solar composition model atmosphere,
  $T_{\rm eff}$ = 40\,000, $\log g=4.0$. At low optical depths,
  ($\tau_{\rm ross} < 10^{-6}$), the total radiative acceleration (dashed line)
  increases and becomes larger than the gravitational acceleration (thin line).
  We apply a numerical cut-off to the radiative acceleration to ensure convergence.
  The assumption of hydrostatic equilibrium breaks down in these superficial layers.
  The arrow indicates our definition of stability against radiation pressure
  ($\Gamma_{\rm rad} < 1$).   \label{GradFig}}
\end{figure}

\begin{figure}
\plotone{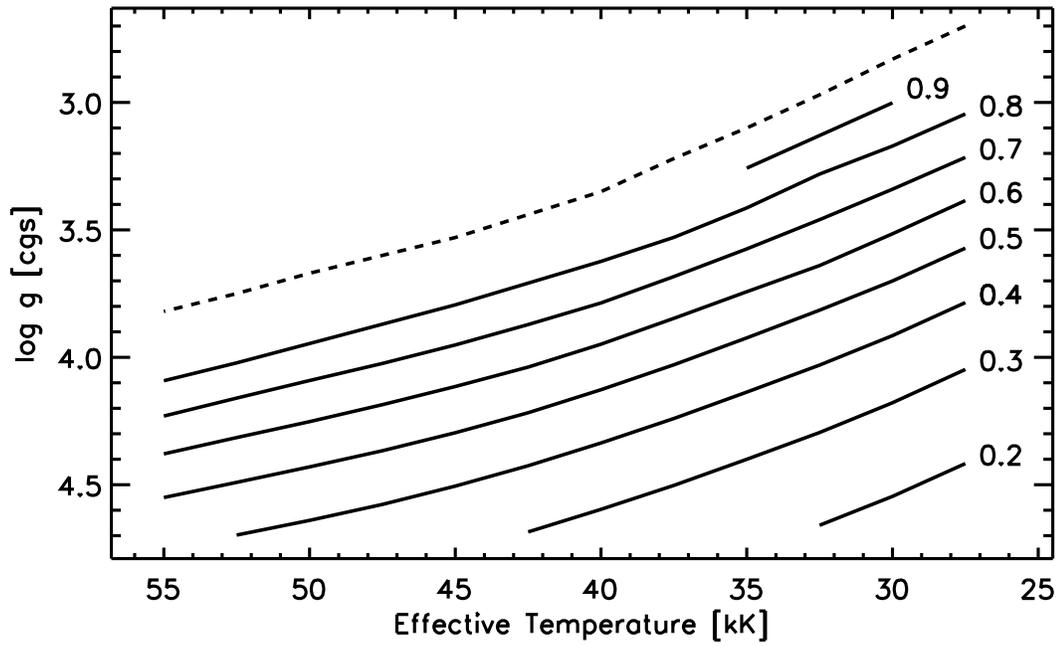}
\figurenum{3}
\caption{ Isolines of radiative acceleration relative to gravitational acceleration,
  $\Gamma_{\rm rad} = g_{\rm rad} / g$, for solar composition model atmospheres,
  as a function of \teff\  and $\log g$. $\Gamma_{\rm rad}$ values are listed right to
  the isolines. The dashed line shows an estimate of the Eddington limit obtained by
  extrapolation. \label{GamaradFig}}
\end{figure}

The lower limit in $\log g$ is set by the Eddington limit. Using NLTE H-He model
atmospheres, we have estimated the lowest surface gravity as a function of \teff\ 
which would correspond to the Eddington limit neglecting the effect of rotation.
This limit is reached for values typically 0.1 to 0.2\,dex smaller than the grid limit.
We present in Fig.~\ref{GradFig} the variation of the radiative acceleration with
depth for a solar composition model, \teff = 40\,000\,K, and $\log g =4.0$.
The radiative acceleration goes through a maximum in the continuum forming
region ($\tau_{\rm Ross}\approx 1$) and shows a strong increase in very superficial
layers ($\tau_{\rm Ross}\leq 10^{-6}$). Near the surface, this model atmosphere
becomes unstable ($g_{\rm rad} > g$). Our modeling assumptions obviously break
down in these layers. We have therefore numerically limited the radiative acceleration
in superficial layers to ensure convergence. This case is quite typical;
a similar behavior is observed in all our model atmospheres, although  higher gravity
models remain numerically stable at every depth (about two thirds of all solar composition
models). The model spectra are not affected by this approximation,
because these superficial layers only influence the strong
resonance lines that form in a stellar wind. The stability of our model atmospheres is
defined by the maximum reached by the radiative acceleration in the photosphere
(see Fig.~\ref{GradFig}). Fig.~\ref{GamaradFig} displays contours of this maximum relative
to the gravitational acceleration, $\Gamma_{\rm rad} = \max(g_{\rm rad})/g$, as a function
of effective temperature and gravity. Models tend to become numerically unstable when
$\Gamma_{\rm rad} > 0.9$, setting thus limits to our grid. Static model atmospheres
are increasingly unrealistic so close to the Eddington limit ($\Gamma_{\rm rad} = 1$).

Finally, we have always adopted a microturbulent velocity, $\xi_{\rm t}=10$\,km/s,
which is a value typically found in spectral analyses of O stars.
These analyses have been performed with static model atmospheres as well as with
wind codes. \citet{JCB346} found that \tlusty\  and CMFGEN models require similar
microturbulent velocities to match the UV metal lines in SMC O dwarfs, showing
therefore that microturbulence is not an artifact of static model atmospheres
but that it indicates the presence of additional non-thermal motions in the atmosphere.
We have included the resulting turbulent pressure term in the hydrostatic
equilibrium equation.

\subsection{Output Products and Availability}

The model atmospheres are available at {\tt http://tlusty.gsfc.nas.gov}.
Each model is characterized by a unique filename describing the parameters
of the model, for example G35000g400v10. The first letter indicates the
composition (see Table~\ref{KeyTbl}), followed by the effective temperature,
the gravity and the turbulent velocity. Each model comes as a set of six files,
with an identical filename's root but a different extension. A complete description
of the files' content and format can be found in \tlusty\  User's guide (see
\tlusty\  web site); for reference, we describe them only very briefly here: \\ [1.5mm]
{\tt model.5}: General input data; \\
{\tt model.nst}: Optional keywords; \\[1.5mm]
{\tt model.7}: Model atmosphere: Temperature, electron density, total density
                and NLTE populations as function of depth; \\[1.5mm]
{\tt model.11}: Model atmosphere summary; \\
{\tt model.12}: Model atmosphere: Similar to {\tt model.7}, but NLTE populations
                 are replaced by $b$-factors; \\[1.5mm]
{\tt model.flux}: Model flux distribution from the soft X-ray to the far-infrared
                    given as the Eddington flux\footnote{The flux at
                    the stellar surface is $F_\nu = 4\pi\,H_\nu$.}
                    $H_\nu$ [in erg\,s$^{-1}$\,cm$^{-2}$\,Hz$^{-1}$] as function of
                    frequency.
                    
The model fluxes are provided at all frequency points included in the calculations
(about 180\,000 to 200\,000 points with an irregular sampling). Additionally,
we have computed detailed emergent spectra with \synspec, version~45, in the
far-UV ($\lambda\lambda$900-2000\,\AA) and in the optical
($\lambda\lambda$3000-7500\,\AA). Filename extensions are {\tt model.uv.7},
{\tt model.vis.7}, respectively, and {\tt *.17} for the continuum spectra.
Only the optical spectra have been computed for the metal-free model atmospheres
({\tt model.hhe.7, model.hhe.17}). Additional spectra in other wavelength ranges,
with altered chemical compositions, or with different values of the microturbulent
velocity can be readily computed using \synspec. This
requires three input files, {\tt model.5, model.7,} and {\tt model.nst},
and the necessary atomic data files (model atoms and the relevant linelist).

\subsection{Interpolation in the Grid}

A grid of model atmospheres is a good starting point for stellar spectroscopic analyses.
We can use it to study trends in spectral features with stellar parameters. However,
the grid sampling is limited by the total number of model atmospheres that can be
computed in some finite period of time. Therefore, we often have to resort to interpolate
in the grid for analyzing individual stars. Since radiative transfer and the complete
model atmosphere problem are highly nonlinear problems, interpolation will result in some
errors. We need to estimate these errors and find the best way to interpolate in the
grid. Here, we consider two possibilities: 
{\sl (i)} we interpolate the model atmosphere (i.e. the temperature and density stratification,
as well as the NLTE populations), and then we recompute a spectrum with \synspec, or
{\sl (ii)} we interpolate directly the detailed spectra, weighted by (\teff$^i$/\teff$^0$)$^4$.
The four models and spectra are otherwise given the same weight.

The \grid\  grid is already finely sampled. We decided therefore to estimate the
interpolation errors by comparing the ``exact'' UV and optical spectra with
spectra interpolated following the two outlined procedures. We have performed this
test on three models, S32500g350v10, L42500g400v10, and G50000g425v10.
We use the four neighboring models ($\pm 2\,500$\,K; $\pm 0.25$\,dex) to obtain the
interpolated spectra. In practice, the interpolated model will always be closer to a
grid point than in these test cases which might therefore be viewed as worst cases.

The two interpolations provide spectra almost identical to the ``exact'' spectra.
The full, detailed comparison requires too much space to appear here, but the figures
are available at the \tlusty\  web site. Minor differences are still found. For example,
the spectrum interpolation results in a few lines being predicted too strong, especially
in the S32500g350v10 case. These lines are much stronger in the coolest model spectrum
due to lower ionization, and this demonstrates the limit of linear spectrum interpolation.
On the other hand, we found that the model atmosphere interpolation results in too strong
\ion{N}{3}\,$\lambda$4640 emissions in the L42500g400v10 models. These emission lines
are very sensitive to the exact $b$-factor ratios, and small differences related to
interpolations may result in larger spectral changes. Based on these limited tests,
we may however conclude that the two interpolation procedures appear both reasonably safe
because of the sufficiently dense sampling of the \grid\  grid. Nevertheless, we feel
that model atmosphere interpolation followed by recomputing the spectrum with \synspec\
remains the safer choice.

The same helium abundance, solar-scaled compositions, and microturbulent velocity
have been adopted for all \grid\  models. \synspec\  spectra can nevertheless be recomputed
with different abundances or microturbulence, using a model atmosphere with the appropriate
stellar parameters. As long as these new values are not changing drastically the overall
opacity and thus potentially changing the atmospheric structure, this procedure should
result in realistic spectra.


\section{Representative Results}
\label{ResuSect}

In this section we show a few representative results. Our aim here
is not to provide a comprehensive display of various models,
but instead to show basic trends of important model parameters 
(temperature structure, emergent spectra) with
effective temperature, surface gravity, and metallicity.

\begin{figure}
\plotone{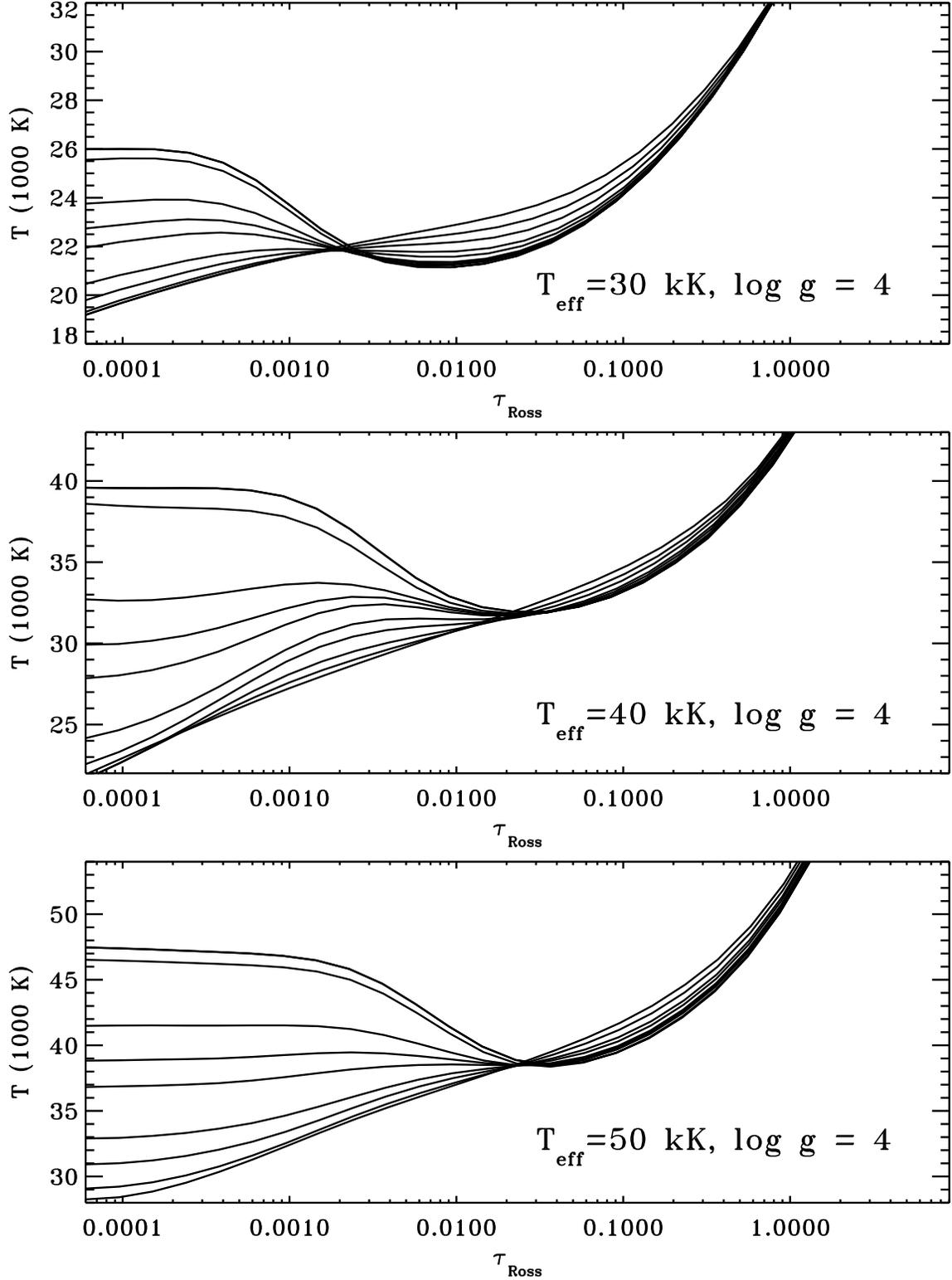}
\figurenum{4}
\caption{ Model atmospheres for
  $T_{\rm eff}$ = 30\,000, 40\,000 and 50\,000\,K (top to bottom panels), 
   $\log g=4.0$, and for various metallicities. At low optical
depths ($\tau_{\rm ross} < 10^{-3}$), the top curves is for a
H-He model, and the temperature is progressively lower when increasing
the metallicity, while reverse is true at deeper layers
($\tau_{\rm ross} > 10^{-2}$). \label{TempFig}}
\end{figure}

Fig.~\ref{TempFig} shows the local temperature for a constant
surface gravity ($\log g = 4$), for three values of \teff\ 
(30\,000, 40\,000 and 50\,000\,K), and for all values of metallicity. 
The temperature
distribution nicely illustrates the basic features of line blanketing,
namely the backwarming -- line blanketing leads to a heating of an atmosphere 
roughly between Rosseland optical depth 0.01 and 1; and the surface
cooling. The zero- and very low-metallicity models exhibit a temperature
rise at the surface, a typical NLTE effect (an indirect heating
effect of hydrogen Lyman and Balmer lines), discovered and
explained already in early days of NLTE model atmospheres by \citet{TNL69}.
This effect competes with surface cooling
due to metal lines. At 50\,000\,K, both effects nearly cancel at 
metallicity 1/50 of solar, while at lower \teff\  the rise
is wiped out at higher metallicities (around 1/10 of solar).
Interestingly, the temperature curves for all metallicities cross
in a narrow range of optical depths; this crossing depth decreases
with decreasing \teff\ 
(being about 0.025 at 50\,000\,K; 0.02 at 40\,000\,K, and 0.002 at 30\,000\,K).

Figures 5-9 display the ionization fractions of all explicit
species, along with LTE fractions (for a direct comparison, 
the LTE fractions are computed
by the Saha formula for temperature and density distributions of
the NLTE models -- that is, we did not recompute self-consistent LTE
model structures). The behavior
of individual species is different, but we can see a
general trend: the fractions of dominant ions 
are usually not sensitive to NLTE effects (that is, they are close
to unity in both LTE and NLTE). The fractions of ions with
lower ionization degrees  are typically
lower in NLTE, because these ions are overionized by the strong radiation
field that originates in deep, very hot layers (notice that the
recombination rate is given through the local temperature). Conversely,
higher ionization degrees are typically larger in NLTE. This ionization
shift is one of the most important NLTE effects in hot star
atmospheres.

Next, we study a sensitivity of emergent flux on $T_{\rm eff}$,
$\log g$, and metallicity. For simplicity, we display spectra computed
directly by  \tlusty\  (files {\tt model.flux} described earlier), 
not detailed spectra computed by \synspec. For clarity, the spectra
are smoothed over 500 frequency points, which roughly simulates a 
10\,\AA~resolution of \citet{CD13} model spectra.

Fig.~\ref{figr1} shows a sequence of model UV spectra with $\log g =4$
and solar metallicity, for various effective temperatures. The Lyman
jump gradually weakens with increasing effective temperature, 
and essentially disappears at 50\,000\,K. Similarly,
one can easily observe gradual weakening and ultimate disappearance of
the \ion{Si}{4}\,$\lambda$1400, \ion{C}{3}\,$\lambda$1176, 977, Ly\,$\alpha$,
etc., features.
Fig. \ref{figr2} shows the effect of surface gravity for a given
effective temperature (40\,000\,K) and metallicity (solar). The most
striking feature is the magnitude of the Lyman jump, which is very
strong at $\log g = 4.5$, and almost disappears at $\log g = 3.5$.
Similarly, the line strengths vary significantly.
Next, we show the effect of metallicity. Fig.~\ref{figr4}
displays models for $T_{\rm eff} = 40\,000$\,K, $\log g = 4$, and
metallicities $Z/Z_\sun$ = 2, 1, 0.5, 0.3, and 0.1. In the
low-metallicity models, the Lyman continuum flux and the
near-UV continuum flux ($\lambda > 1700$\,\AA) is lower than that for 
higher metallicities. This relation is reversed in the region
crowded with metal lines, $900\leq\lambda\leq 1700$\,\AA.

Finally, we compare our model fluxes to \citet{CD13} models. Because of
different averaging procedures used in both sets of model spectra, 
the comparison
is not exact, nevertheless shows the general effect. We compare
three models for $(T_{\rm eff}, \log g)$ pairs equal to
(40\,000, 4.5), (35\,000, 4.0), and (30\,000, 4.0), and for solar
metallicity -- Fig.\ref{figk2}.
The agreement between our and Kurucz models is very good in the
optical spectrum for all three models (lower panel). 
For $T_{\rm eff} = 30\,000$ K, the
agreement between \grid\  and Kurucz fluxes is rather good in the
UV spectrum ($\lambda > 900$\,\AA), while for the hotter
models our NLTE flux is lower. In contrast, NLTE models predict
higher flux than Kurucz model in the Lyman continuum, although
the effect is relatively small at 40\,000\,K (see also \S\,\ref{BCSect}).

This exercise shows that using Kurucz model is still a reasonable
choice for analyzing {\em low-resolution} UV and optical spectra of O
stars. It should be realized, however, that the full strength
of NLTE models lie in their ability to provide reliable high-resolution
spectra and profiles of individual lines that can be used
for determining the stellar parameters and the surface chemical composition.
We will study various systematic effects and differences between LTE and NLTE models,
and their influence on deduced stellar parameters, in a future paper.


\section{Bolometric Corrections and Ionizing Fluxes}
\label{BCSect}

\subsection{Bolometric Corrections}

Bolometric corrections (Table~\ref{BCTbl}) are calculated using the following expression:
\begin{equation}
BC = m_{\rm bol} - V = (-2.5\log F_{\rm bol} - 11.487) - (-2.5\log F_V - 21.100)
\end{equation}
where $F_{\rm bol}$ is the bolometric flux and $F_V$ is the flux through
the Johnson $V$ filter, computed from the model atmospheres' flux distribution.
The bolometric flux is computed by trapezoidal integration over the complete frequency
range, while we have used an IDL version of the program {\sc ubvbuser} \citep{CD13}
kindly made available by Wayne Landsman to obtain the $V$ magnitudes.  The first constant
is defined by assuming a solar constant, $f_{\rm bol}^\sun = 1.371\,10^6$\,erg\,s$^{-1}$\,cm$^{-2}$,
a solar visual magnitude, $V_\sun = -26.76$, and a solar bolometric correction, $BC_\sun = -0.07$,
while the second constant defines the zero point of the $V$ magnitude scale \citep{BCP98}.

Fig.~\ref{BCFig} illustrates the major dependence of the bolometric correction with
the effective temperature. Over the parameter space of the grid, a multi-linear
regression yields the relation:
\begin{equation}
\label{BCEq}
BC = 27.43 - 6.78\times\log(T_{\rm eff}) + 0.06\times Z/Z_\sun
\end{equation}
An inspection of Table~\ref{BCTbl} shows that the dependence with gravity is generally even
smaller than the metallicity term. We have therefore neglected the gravity dependence and
Eq.~\ref{BCEq} has been established with values from $\log g=4.0$ models.
The typical accuracy of bolometric corrections derived from Eq.~\ref{BCEq} is $\pm$0.05\,mag.
Larger differences up to 0.1\,mag between the models and Eq.~\ref{BCEq} values are found
for the coolest models.

We compare our bolometric corrections to values calculated for \citet{CD13} models and
extracted from \citet{BCP98}. We use the same zero point for bolometric corrections.
We find almost identical values for solar composition
models with $\log g = 4.0$, and 4.5. The largest differences reach 0.03\,mag at 
\teff\,=\,47500\,K, and 0.05\,mag at \teff\,=\,27500\,K, adding thus to the good agreement
found in the previous section between spectral energy distributions (Fig.~\ref{figk2}).

Our relation (Eq.~\ref{BCEq}) is close to the Vacca, et al. (1996) calibration,
which is based on a number of diverse NLTE analyses.
Their bolometric corrections are systematically smaller by 0.12~mag. Our relation
therefore yield stellar luminosities 12\% fainter than previously estimated from
\citet{Vacca96}.

\subsection{Ionizing Fluxes}

We have calculated the ionizing fluxes in the hydrogen Lyman and in the
\ion{He}{1}\,$\lambda$504 continua. They are listed in Tables~\ref{Q0Tbl} and~\ref{Q1Tbl}.
The ionizing fluxes, $q_0$ and $q_1$, are given as logarithms of the number of photons
in these two continua, per second and per square centimeter at the stellar surface.
We do not provide the ionizing flux $q_2$ in the \ion{He}{2} Lyman continuum.
\citet{Gab91} indeed showed that the \ion{He}{2} continuum is formed in a stellar wind
(due to high opacity in this continuum), and that spherical extension and shocks are
necessary model ingredients to predict fluxes that are in rough agreement with observed fluxes.
However, the contribution of the \ion{He}{2} Lyman continuum to $q_0$ and $q_1$ remains
very small. Any error in $q_2$ results therefore in unsignificant changes
of $q_0$ and $q_1$. The latter fluxes are mostly sensitive to a correct treatment
of NLTE metal line blanketing, which in turn is crucial
to predict the hydrogen and helium ionization
structure and the photospheric temperature stratification reliably.

The dependence of the ionization fluxes with \teff, $\log g$, and $Z/Z_\sun$, is displayed
in Fig.~\ref{QQFig}. Lower gravity models have larger ionizing fluxes due to higher
hydrogen and helium ionization (thus the Lyman and $\lambda$504 jumps are smaller).
Model atmospheres with low metallicity have lower Lyman continuum fluxes. The behavior
of the \ion{He}{1} continuum fluxes is more complicated: at low \teff, metal-poor model
atmospheres have markedly lower $q_1$ fluxes, while at \teff\  higher than about 33\,000\,K,
they have slightly larger $q_1$ fluxes. Lower metal abundances reduce the blanketing effect,
and thus result in lower temperatures in the continuum-forming layers and lower fluxes.
The higher $q_1$ fluxes in hot, metal-poor models come from smaller metal line opacities
in this spectral range.

We have compared our ionizing fluxes to other recent works.  \citet{Vacca96} used
\citet{CD13} LTE model fluxes to compute $q_0$ and $q_1$. They list 43 values for parameters
covering the domain of O stars. We interpolate bilinearly in $\log T_{\rm eff}$ and
$\log g$ in Tables~\ref{Q0Tbl} and~\ref{Q1Tbl} and compare to Vacca et al.'s values.
The $q_0$ values agree very well, with no systematic differences; we still found differences
up to 0.1-0.2\,dex for some parameters (mostly the coolest models). On the other hand,
Vacca et al.'s $q_1$ values are on average 0.25\,dex smaller than ours; individual differences
range typically from 0.1 to 0.5\,dex, increasing towards cooler effective temperatures.
These differences are quite significant for nebular studies.

We have also compared our results to values derived from NLTE, line-blanketed, expanding model
atmospheres, calculated with program {\sc wm-basic} \citep{Mun01}. The $q_0$ values agree
also very well, although our values are slightly larger. The values agree within 0.1\,dex,
with the exception of the two models, D-30 and S-30, at \teff = 30\,000\,K, where {\sc wm-basic}
models predict ionizing fluxes smaller by 0.57 and 0.24\,dex, respectively.


\section{Model Sensitivity to Atomic Model Atoms}
\label{ErrSect}

We have aimed at incorporating all important opacity sources and all relevant
atomic processes that determine the excitation and ionization balance in the model
atmospheres. To investigate the sensitivity of the model atmospheres to our handling
of atomic data, we present here a few test cases where we adopted different
model atoms. We consider here two potential issues: {\sl (i)} the size of the iron
model atoms and the representation of the line opacity; {\sl (ii)} the omission of
highly-excited levels in light elements (mostly, CNO).

\citet{CD22} lists thousands of energy levels for every iron ion. How should we group
them? Our model atoms group them into 30 to 50 superlevels per ion. On the other hand, 
\citet{DD99} have grouped all individual levels into 7 to 9 NLTE superlevels
for each iron ion. We have compared the effect of different Fe model data on the resulting
model atmospheres, with a test case, $T_{\rm eff} = 35\,000$\,K, $\log g = 4.0$,
solar composition, and $\xi_{\rm t}=10$\, km/s. The reference model is extracted from the
\grid\  grid. The first three models deal with the representation of line opacity: OS with larger
frequency step, ODF, and line strength selection (Fe and Ni lines are dynamically selected
depending on a strength criterion, see \S\,\ref{iron}).
In the case of the ODF model (Gb), all the Fe and Ni lines
are used to set up the ODFs. In the second set of models, we have built different
Fe~{\sc iv} and Fe~{\sc v} model atoms (they are the dominant ions, see Fig.~\ref{IonEFig}).
All individual levels are included in models Gd and Ge, but grouped in more
(74\&69 {\sl vs.\/} 43\&42) or less (33\&19) superlevels.  In the last model, all the Fe and Ni
levels above the ionization limits are omitted. Table~\ref{TestTbl} summarizes the
properties of the different models.

Fig.~\ref{FeTFig} illustrates the changes in the model temperature atmosphere structure due
to different ways of including the iron data. This is a condensed way to assess differences
between model atmospheres. We may indeed expect that model spectra change with
changes in the temperature structure (this does not exclude of course that some line profiles
might be different even with very close temperature structures, see below). First, differences
are larger with ODF's than when adopting a rather large sampling step. We interpret this as
reflecting incorrect blends between ODFs and important lines of light elements. The larger
sampling step is probably appropriate in many cases ($\Delta T\approx 100$\,K), and reduces
the computing time by a factor of about three (the reduction in the number of frequency points).
The largest difference arising due to the line strength selection criterion is seen at depth,
because of a large change in the number of selected \ion{Fe}{6} lines. However, this
does not affect the predicted spectrum. Differences in the layers where the spectrum
is formed are small ($\Delta T < 100$\,K) showing that most of the line-blanketing
effect is due to the strongest 10$^4$ to 10$^5$ lines rather than to millions of
weaker lines. The lower panels of Fig.~\ref{FeTFig} show the results related to the model
atoms. First, we see that increasing the number of superlevels results in small changes,
while the model with less superlevels shows a larger systematic temperature difference.
Although limited in scope (one set of stellar parameters), this test thus supports
a claim that we have adopted Fe model atoms of reasonable sizes for our grid of
model atmospheres. Finally, the levels above the ionization limits have a minimal
effect on the model atmosphere structure.

In a second test, we have investigated the omission of highly-excited levels of light
elements. Sect.~\ref{AtomSect} details the adopted model atoms. In some cases, levels
with main quantum number, $n\geq 7$, have been neglected (e.g., \ion{C}{3} and \ion{N}{4}).
We have built a new \ion{C}{3} model atom which includes all levels below the ionization
limit that are considered in OP calculations. The new model atom includes the first 34
levels as individual explicit levels and group higher levels into 12 superlevels.
We have recomputed one model atmosphere, S35000g325v10, assuming the extended \ion{C}{3}
model atom. The atmospheric structure is changed very little ($\Delta T < 8$\,K), and
this omission is therefore mostly of no consequence.

However, the total recombination rate, from C$^{+3}$ to C$^{+2}$ is significantly
increased, by a factor 2 to 3. Recombination lines in the optical, e.g.
\ion{C}{3}\,$\lambda$4650, show thus significantly stronger emission in the new model.
We note here in passing that these ``Of-type" lines are predicted in emission by
plane-parallel, hydrostatic, NLTE model atmospheres at low gravities \citep{MH73}. These are
NLTE recombination lines and emission is neither a consequence of a stellar wind nor of an
extended atmosphere. We caution therefore against using the current grid for quantitative
analysis of weak absorption or emission features in the optical spectrum without a
careful investigation case by case. A future update is expected.


\section{Discussion and Conclusions}
\label{ConclSect}

We have constructed a comprehensive grid of metal line-blanketed, NLTE,
plane-parallel, hydrostatic model
atmospheres for the basic parameters appropriate to O-type stars.
The \grid\  grid considers 12 values of effective temperatures 
$27\,500$\,K $\leq$\teff\ $\leq 55\,000$\,K, with 2\,500\,K steps, 
8 surface gravities, $3.0\leq\log g\leq 4.75$ with  
0.25\,dex steps, and 10 chemical compositions, from metal-rich 
relative to the Sun to metal-free.  The lower limit of
$\log g$ for a given \teff\  is actually set by an approximate 
location of the Eddington limit. 
The selected chemical compositions have been chosen to cover a number
of typical environments of massive stars: the galactic center,
the Magellanic Clouds, Blue Compact Dwarf galaxies like I\,Zw\,18, 
and galaxies at high redshifts.  The paper contains a description
of the \grid\  grid and some illustrative examples and comparisons.
The complete \grid\  grid is available at our website 
at {\tt http://tlusty.gsfc.nasa.gov}.

We are aware that the presented grid cannot represent the last word
in O star model stellar atmospheres.  However, we intended to provide 
a more or less definitive grid of model atmospheres in the context of
one-dimensional, plane-parallel, homogeneous, hydrostatic models
in radiative equilibrium.
We have attempted to take into account essentially all important
opacity sources (lines and continua) of all astrophysically important
ions. Likewise, we have attempted to consider all relevant atomic 
processes that determine the excitation and ionization balance of
all such atoms and ions.  In other words, since the computational
effort behind constructing the above described 690 models is
enormous, we intended to compute as complete and realistic models
as possible, in order to avoid a need to recompute a grid in a near
future should some neglected mechanism or an opacity source is found to
be important.

Although we spent all effort to make sure that the treatment
of atomic physics and opacities is as complete and accurate
as possible, there are still several points we are aware of that 
were crudely approximated.  Those approximations were necessitated
not by shortcomings of our modeling scheme or a lack of adequate
computer resources, but by the present lack of sufficient atomic data.
These approximations include (roughly in the order of estimated
importance):

$\bullet$ A crude and approximate treatment of collision rates for 
forbidden transitions of Fe and Ni. As discussed in \S\,\ref{AtomSect}, 
since no approximate formula or a simple prescription is available, 
we have somewhat arbitrarily adopted the Eissner-Seaton formula with 
$\gamma(T)=0.05$. The IRON project provides, and will keep providing
in future,  theoretical calculations for selected 
transitions for various ions of iron; we plan to implement all the 
newly available collision rates in future calculations in order 
to check their influence explicitly.

$\bullet$ A neglect of very high energy levels of light metals. 
The Opacity Project, which is the basic source of our atomic data,
typically considers energy levels only up to a certain principal
quantum number (usually 10). The higher
levels are neglected. This does not usually cause any significant 
errors in the excitation balance, but in some cases
recombinations to these neglected levels may represent a non-negligible
contribution to the total recombination rate -- the global ionization 
balance may thus be somewhat in error. 

$\bullet$ Inaccuracies in, or a lack of, computed photoionization cross-sections. 
Although the Opacity Project provided a large number of cross-sections, 
many are still missing. We have approximated all cross-sections for 
which we did not find any published data by hydrogenic cross-sections.
However, since we have adopted accurate data for the most
important transitions, we do not believe that this may be a source
of significant errors.

$\bullet$ We did not consider charge exchange reactions. In this case, some
necessary atomic data are available, but in view of high
temperature of the atmospheres of O-type stars, we do not expect
the charge exchange reactions to have any significant contribution
to the computed NLTE excitation/ionization balance.

For completeness, we repeat the list the most important physical 
approximations of our model grid:

$\bullet$ Plane-parallel atmosphere in hydrostatic equilibrium. 
This is perhaps the most important and restrictive assumption. 
Because of its importance, we have discussed it at length in 
\S\,\ref{StaticRelev}. Essentially, we argued
that while the wind features cannot obviously be described
by means of our models, the rest of the UV and optical
spectra is well described, unless the wind is
truly extreme. In any case, the \grid\  grid should not 
be used for analyzing Luminous Blue Variables or Wolf-Rayet
stars, where already the continuum is formed in the wind; and
the grid should be used with caution in the case of Of supergiants.
 
$\bullet$ Homogeneity (i.e., a 1-D atmosphere). This means that any surface 
inhomogeneities are neglected. However, our understanding of 
inhomogeneities in hot star atmospheres is very poor, and at present 
there is little what can be done on a quantitative level.

$\bullet$ Radiative equilibrium. Our approach neglects any source of 
mechanical energy dissipation, shocks, non-thermal X-rays, and
related phenomena. Again, there is no approach at present that
would address those issues from the first principles, and therefore
any modeling of such phenomena is necessarily extremely crude.

Despite of the long list of approximations and uncertainties,
we believe that the \grid\  grid represents a definite improvement
over previous grids of O star model atmospheres.
We believe that we have constructed a more or less definitive grid of 
models in the context of one-dimensional, plane-parallel, homogeneous,
geometry in a hydrostatic and radiative equilibrium, with all
important opacity sources (metal lines) included, and without
any unnecessary numerical approximations.
We hope that the \grid\  grid of models will be useful for a number of 
years to come, until a new-generation models with multi-dimensional,
self-consistent radiation magneto-hydrodynamic description will be 
developed.

\acknowledgments

We are indebted to Sally Heap for her continuing encouragement
and motivation to compute this grid, and for suggesting a number of
various improvements of the modeling strategy. We acknowledge
helpful discussions and collaboration with many colleagues, most
importantly with John Hillier, David Hummer, Dimitri Mihalas,
Klaus Werner, Stefan Dreizler, Stefan Haas, and the late Keith Smith.  
On the computational side, we would like to thank most sincerely
Keith Feggans for his invaluable assistance, 
and for his patience to answer our many calls to help whenever any 
computer-related problem occurred, day or night, during our work 
on computing model atmospheres at NASA/GSFC.
Last but not least, we are greatly indebted to all the physicists
who provided, and are still providing, all the necessary atomic data, in particular
the OP/IP team, Bob Kurucz, and the NIST team. The present sophistication of
current model atmospheres would simply not been achievable without all their
work during many years. This work was supported by NASA grant NAG5-10895
(FUSE B134 program), NASA NRA-01-01-ADP-099 grant, and by several grants (GO 7437,
AR 7985, GO 9054) from the Space Telescope Science Institute, which is operated
by the Association of Universities for Research in Astronomy, Inc., under NASA
contract NAS5-26555.


\clearpage

\clearpage

\clearpage

\clearpage

\clearpage

\begin{figure}
\figurenum{5}
\plotone{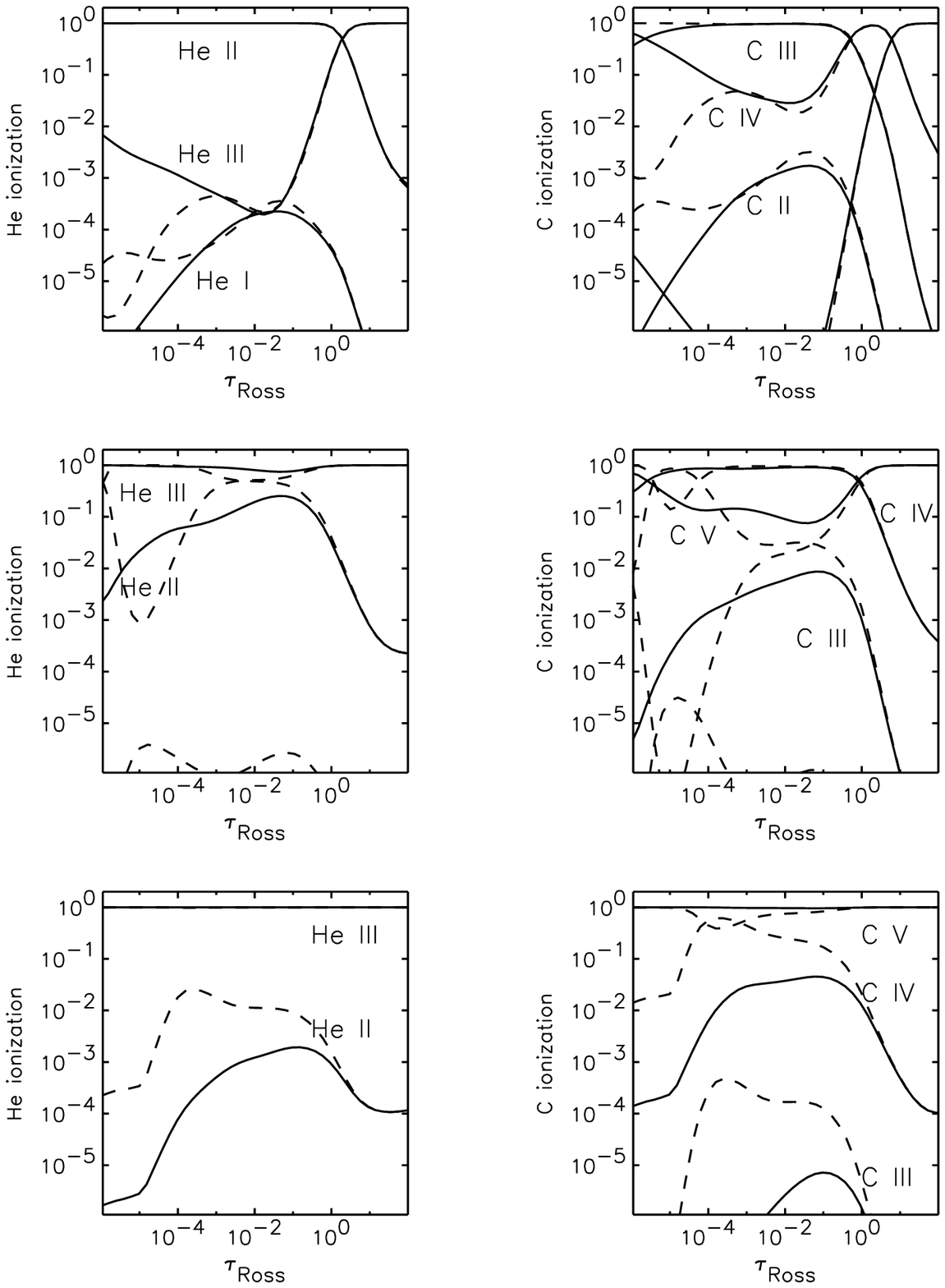}
\caption{Ionization fractions of helium and carbon in three model atmospheres, 
   \teff = 30\,000, 40\,000 and 50\,000\,K (top to bottom panels), 
   $\log g=4.0$, and solar composition. LTE ionization is shown with dashed lines. \label{IonAFig}}
\end{figure}

\clearpage

\begin{figure}
\figurenum{6}
\plotone{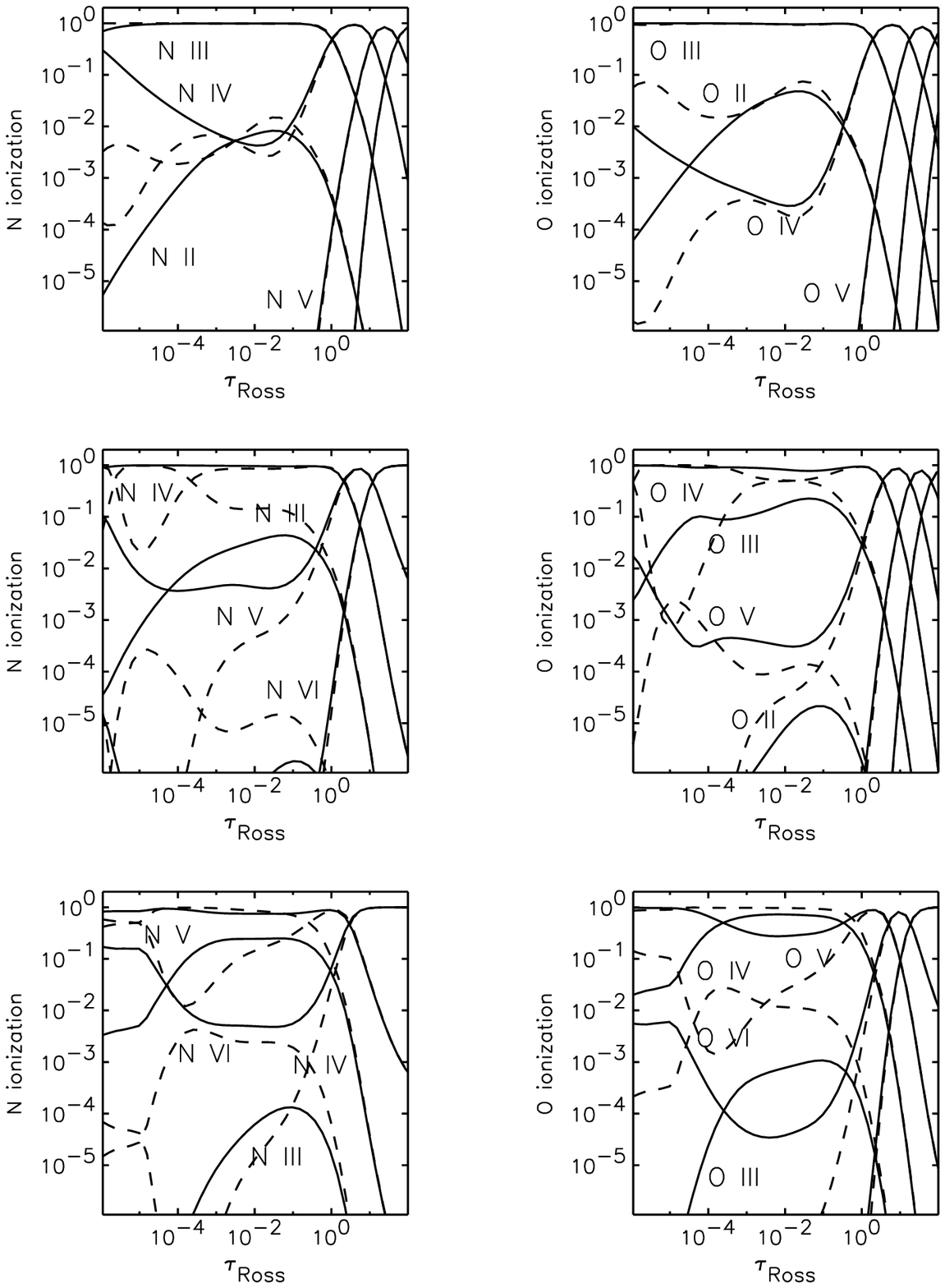}
\caption{Ionization fractions of nitrogen and oxygen in three model atmospheres, 
   \teff = 30\,000, 40\,000 and 50\,000\,K (top to bottom panels), 
   $\log g=4.0$, and solar composition. LTE ionization is shown with dashed lines.  \label{IonBFig}}
\end{figure}

\clearpage

\begin{figure}
\figurenum{7}
\plotone{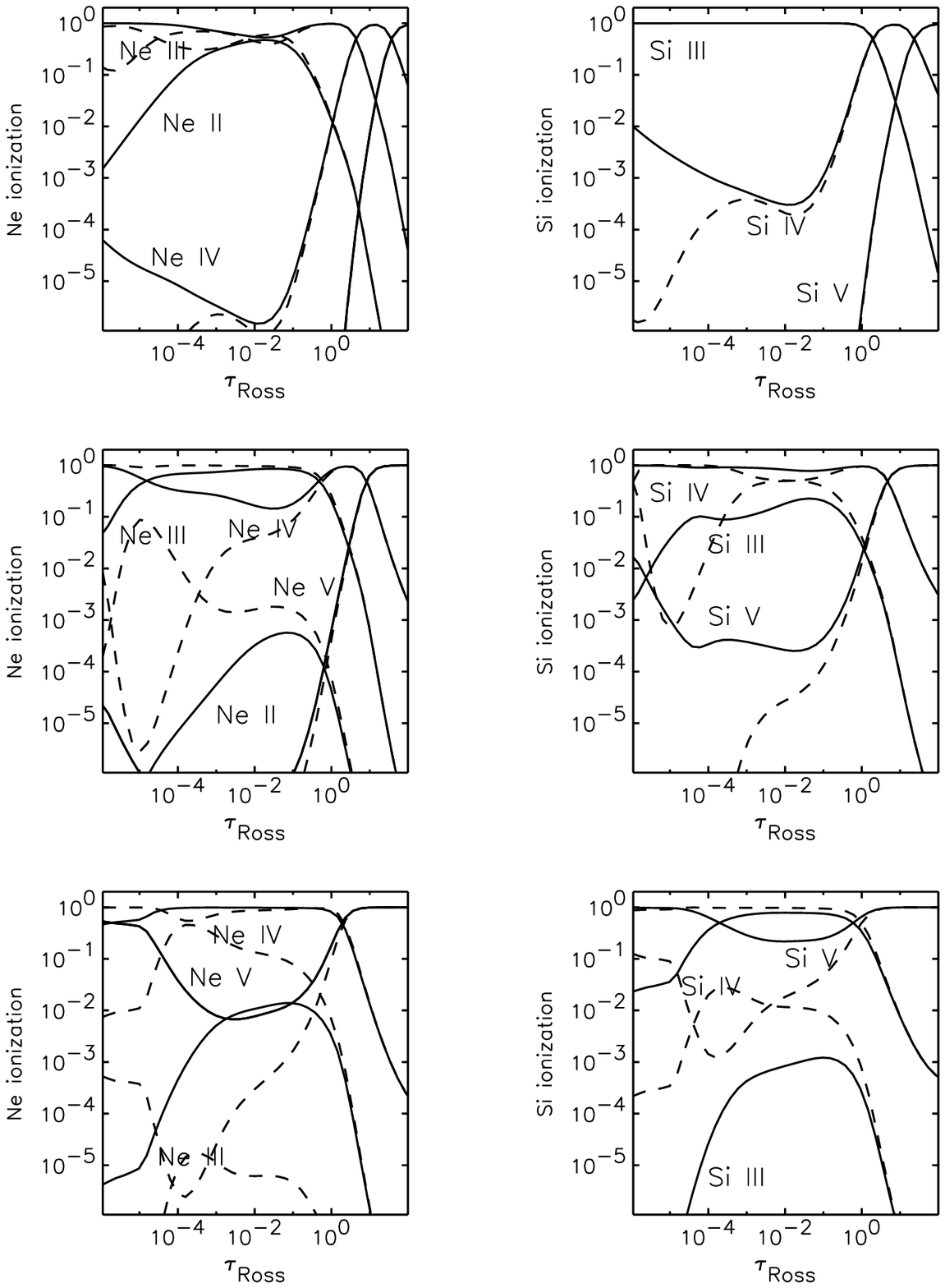}
\caption{Ionization fractions of neon and silicon in three model atmospheres, 
   \teff = 30\,000, 40\,000 and 50\,000\,K (top to bottom panels), 
   $\log g=4.0$, and solar composition. LTE ionization is shown with dashed lines.  \label{IonCFig}}
\end{figure}

\clearpage

\begin{figure}
\figurenum{8}
\plotone{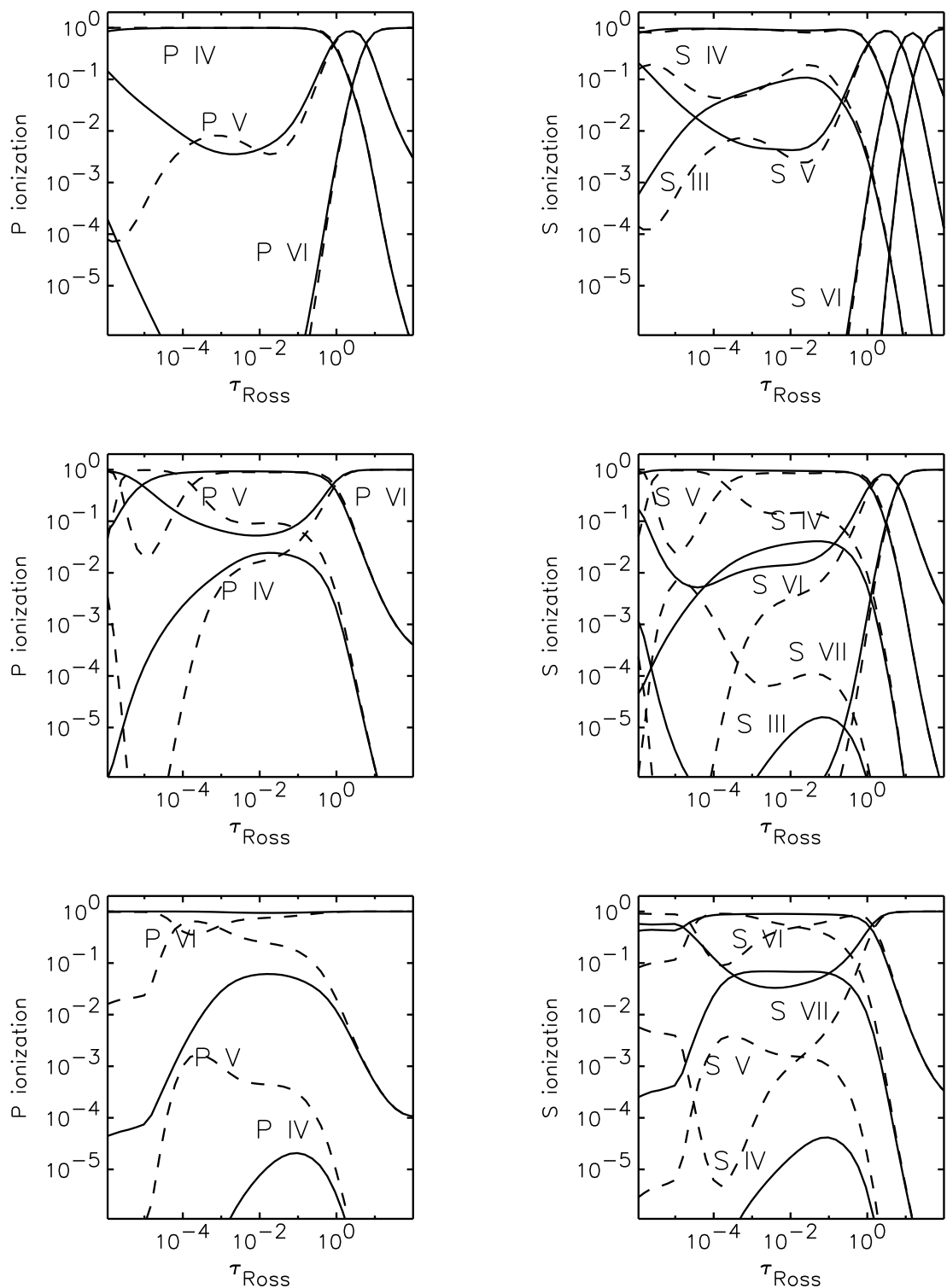}
\caption{Ionization fractions of phosphorus and sulfur in three model atmospheres, 
   \teff = 30\,000, 40\,000 and 50\,000\,K (top to bottom panels), 
   $\log g=4.0$, and solar composition. LTE ionization is shown with dashed lines.  \label{IonDFig}}
\end{figure}

\clearpage

\begin{figure}
\figurenum{9}
\plotone{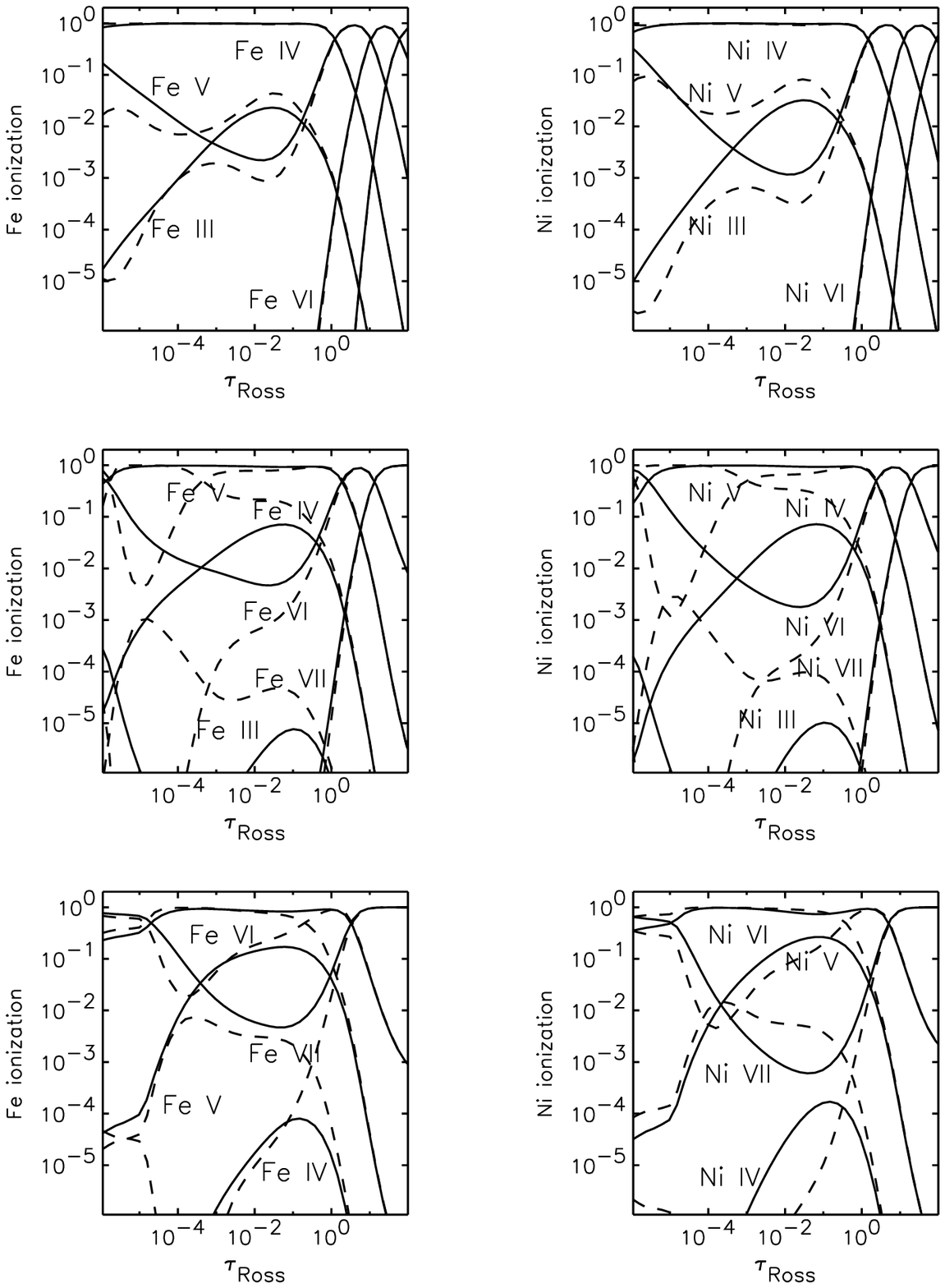}
\caption{Ionization fractions of iron and nickel in three model atmospheres, 
   \teff = 30\,000, 40\,000 and 50\,000\,K (top to bottom panels), 
   $\log g=4.0$, and solar composition. LTE ionization is shown with dashed lines.  \label{IonEFig}}
\end{figure}

\clearpage

\begin{figure}
\figurenum{10}
\plotone{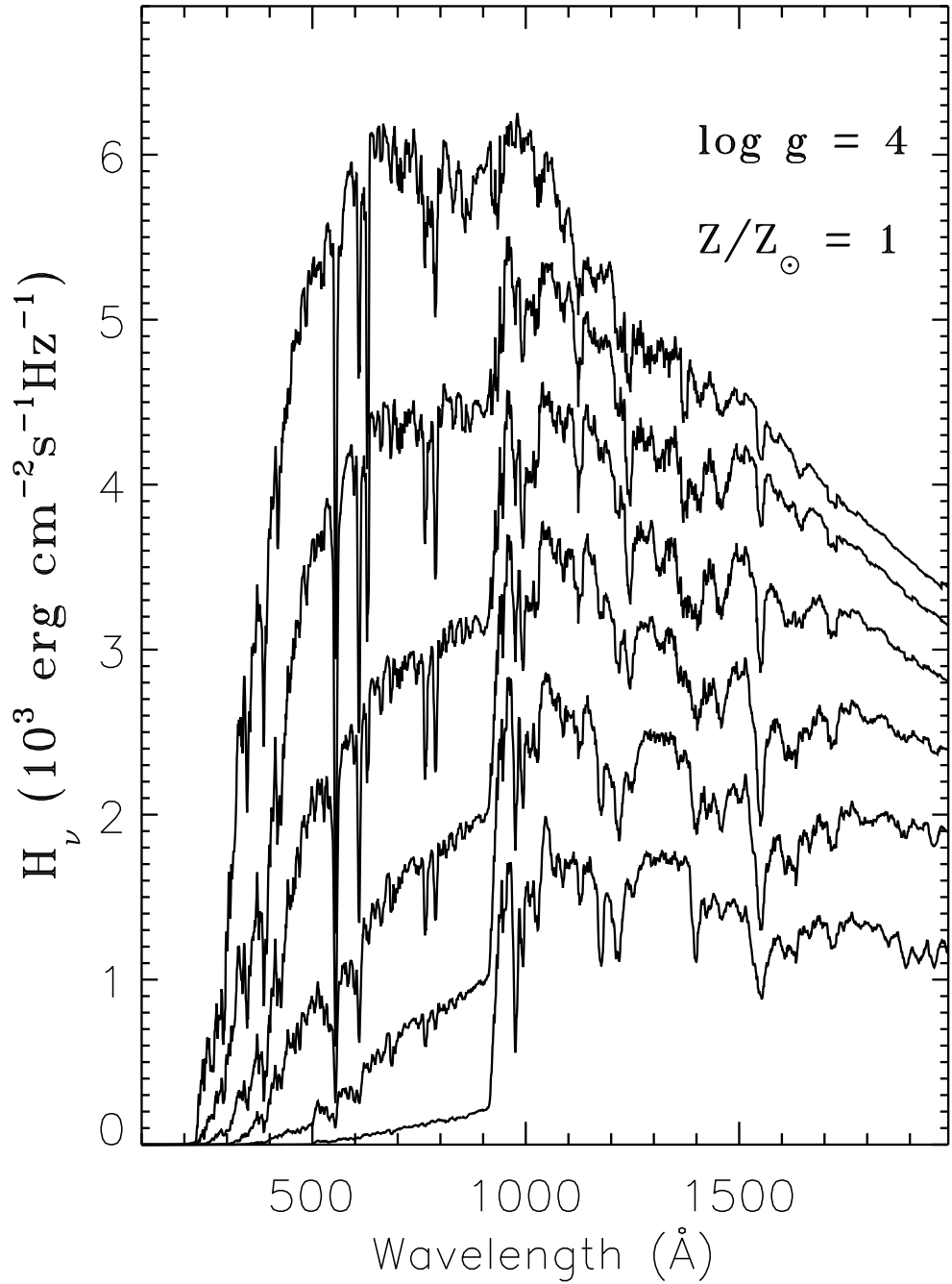}
\caption{ Predicted flux for model atmospheres for $\log g=4.0$, solar composition,
and various effective temperatures  -- from top to bottom 
\teff\  = 55, 50, 45, 40, 35, and 30\,kK. \label{figr1}}
\end{figure}

\clearpage

\begin{figure}
\figurenum{11}
\plotone{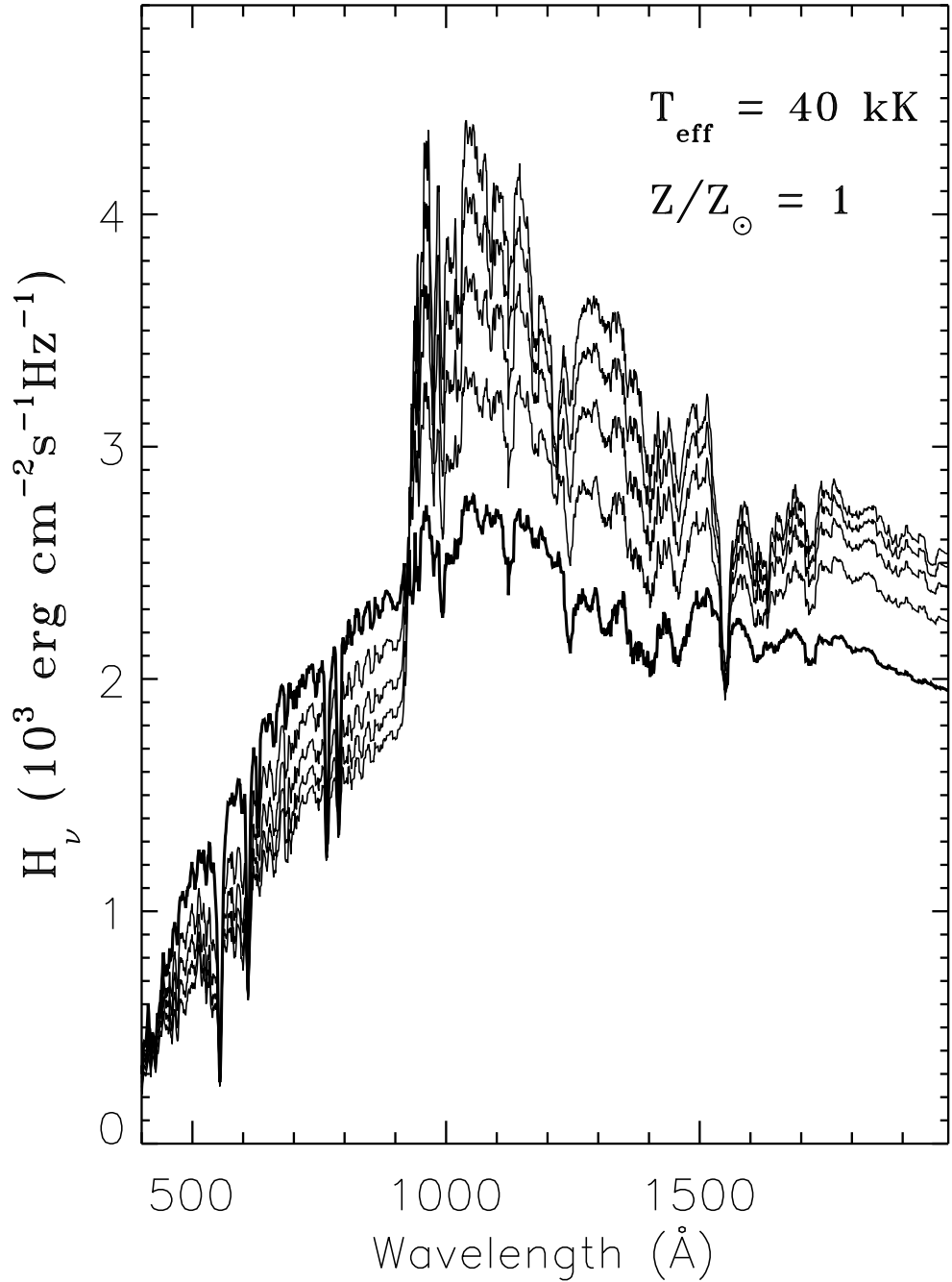}
\caption{ Predicted flux for model atmospheres for
  \teff\  = 40\,000\,K, solar composition, and various surface
gravities  -- from top to bottom  at the Balmer continuum 
($\lambda >912$\,\AA)
$\log g =$ 4.5, 4.25, 4.0, 3.75, and 3.5; the lowest-gravity model
is drawn by bold line. Notice that the dependence of flux on $\log g$
is reversed in the Lyman continuum ($\lambda < 912$\,\AA). \label{figr2}}
\end{figure}

\clearpage

\begin{figure}
\figurenum{12}
\plotone{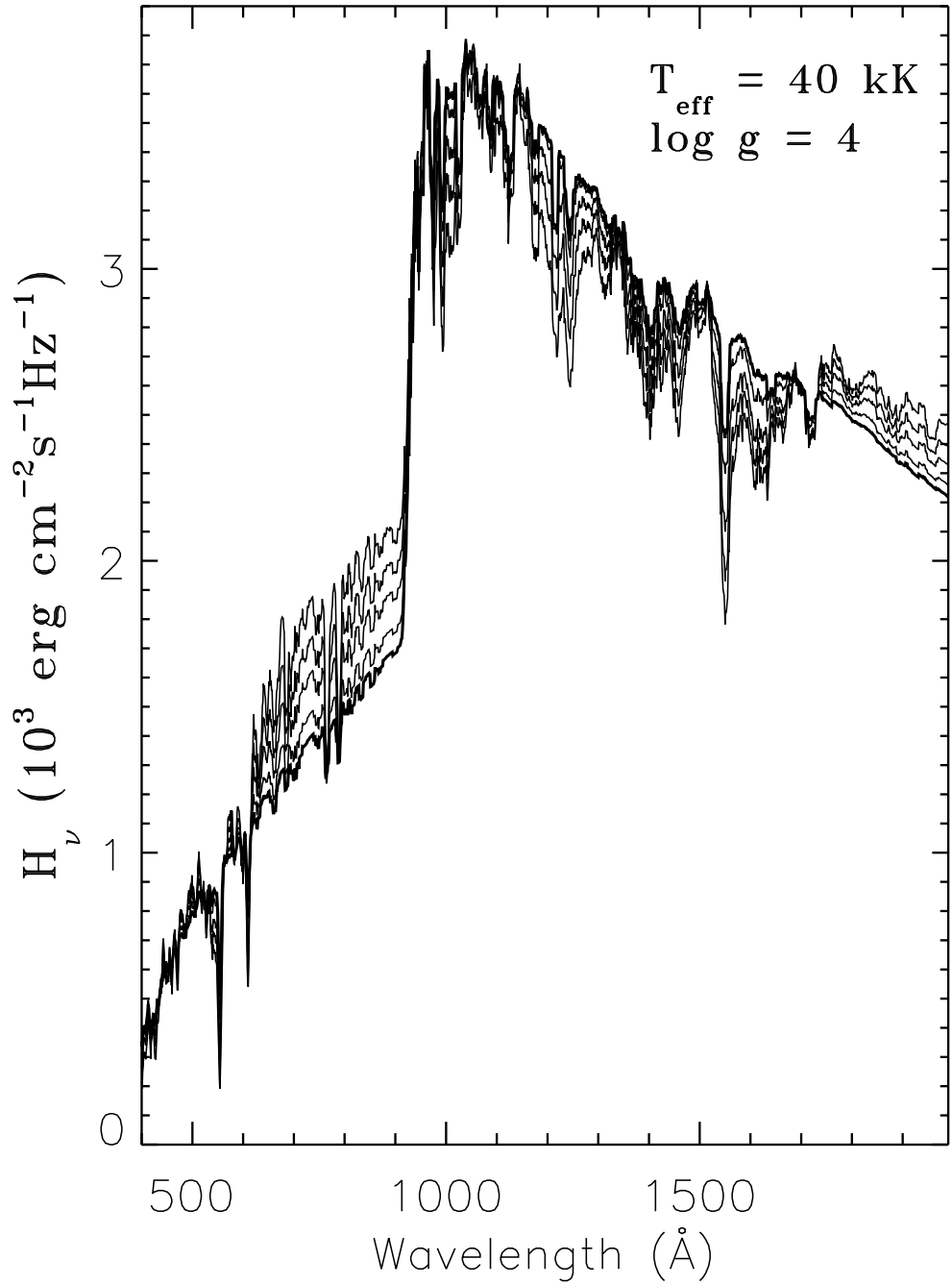}
\caption{ Predicted flux for model atmospheres for
  $T_{\rm eff} =40$ kK, $\log g=4.0$, and various metallicities --
$Z/Z_\sun = 2, 1, 1/2, 1/5, 1/10$. The flux for the lowest-metallicity
model is drawn by a bold line. The flux generally decreases with
decreasing metallicity for 
$\lambda < 900$ \AA, and $\lambda > 1700$ \AA,
while the relation is reversed for $ 900 < \lambda < 1700$ \AA. \label{figr4}}
\end{figure}

\clearpage

\begin{figure}
\figurenum{13}
\plotone{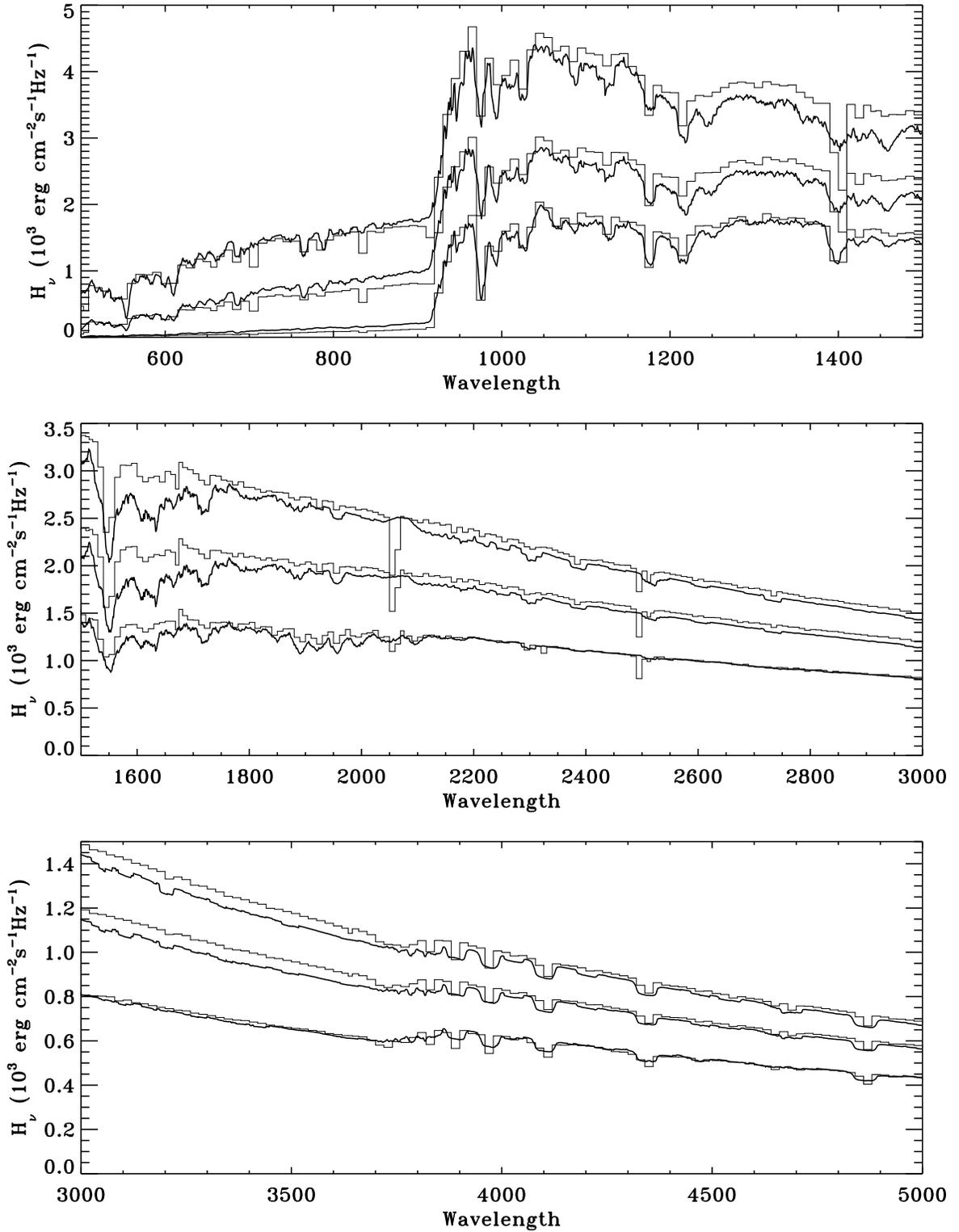}
\caption{ Predicted flux for model atmospheres for three model 
atmospheres with $(T_{\rm eff}, \log g)$ equal to (40000\,K, 4.5);
(35000\,K, 4.0), and (30000\,K, 4.0) -- bold lines; compared to Kurucz 
models with the same parameters -- thin histograms. 
Different panels show different spectrum regions. \label{figk2}}
\end{figure}

\clearpage

\begin{figure}
\figurenum{14}
\plotone{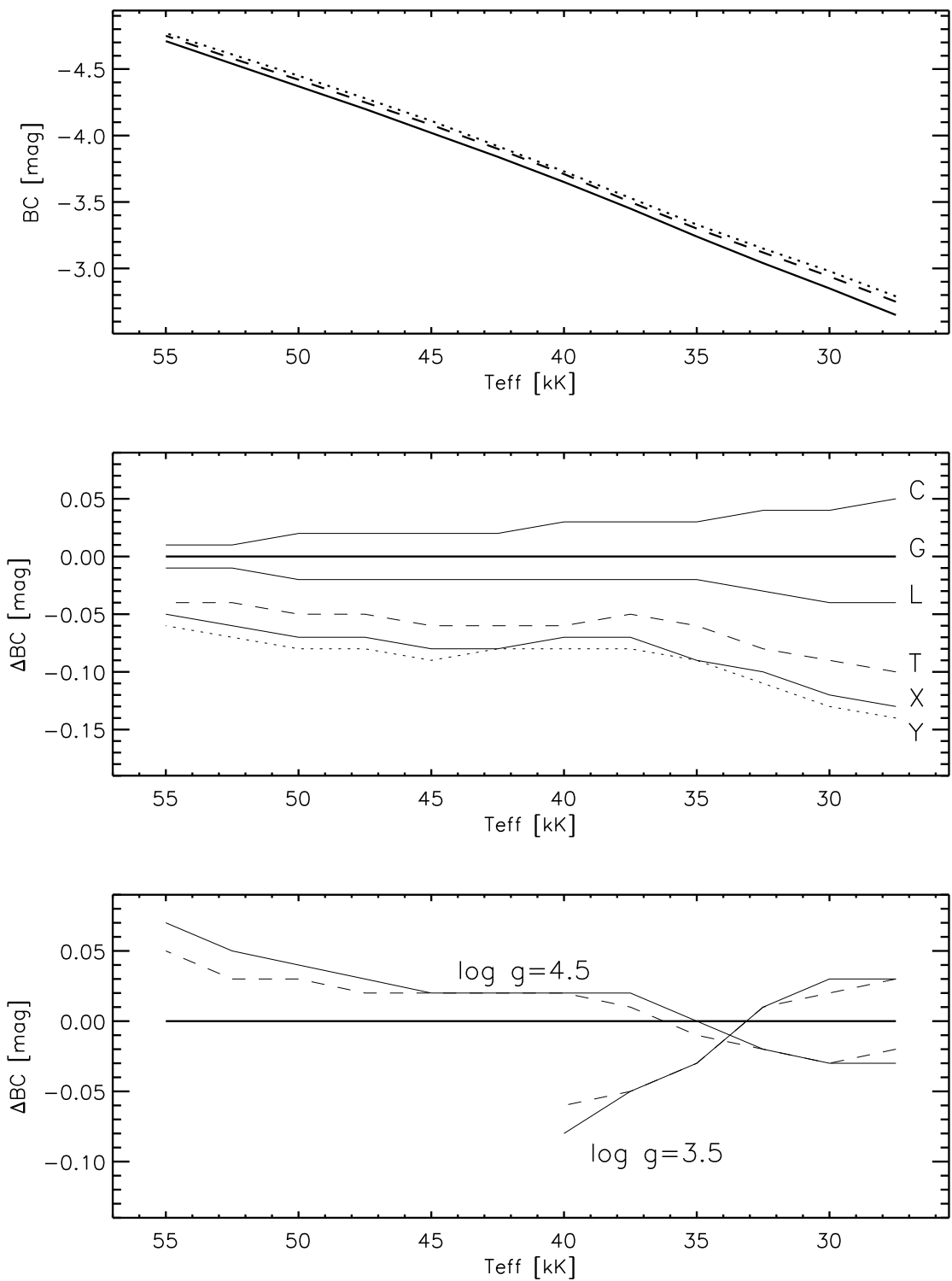}
\caption{Bolometric corrections {\it vs.} effective temperature. The top panel
displays the relation for $\log g=4.0$, solar composition (full), 1/10 (dashed) and
1/1000 solar (dotted). The middle panel expands on the effect of metallicity on BC,
displaying differences relative to solar models ($\log g=4.0$). The bottom panel
shows the effect of different gravities (full: solar; dashed: 1/10 solar).  \label{BCFig}}
\end{figure}

\clearpage

\begin{figure}
\figurenum{15}
\plotone{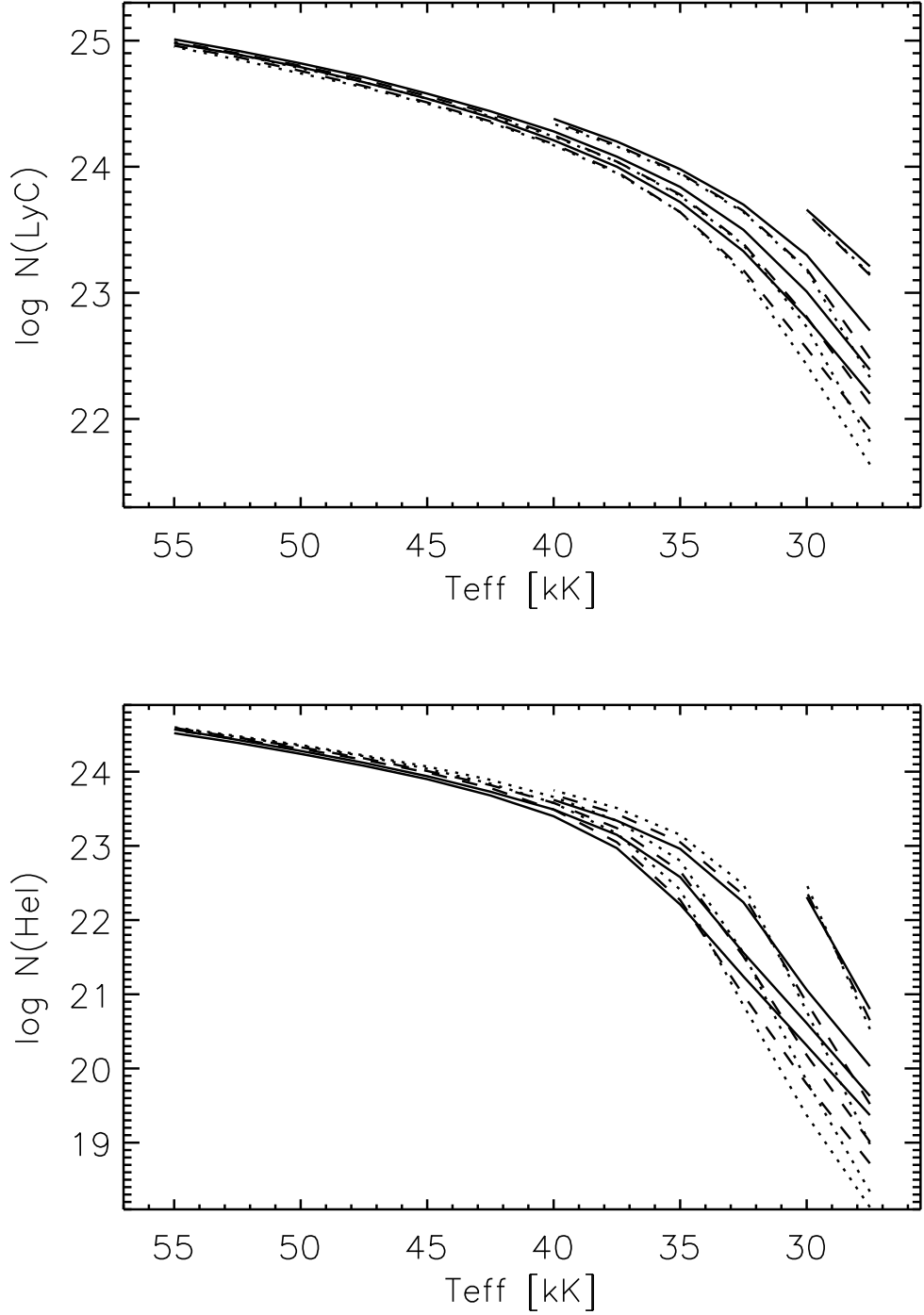}
\caption{Ionization fluxes [s$^{-1}$ cm$^{-2}$] {\it vs.} effective temperature.
Top: log of number of Lyman continuum photons; bottom: log of number of \ion{He}{1}$\lambda$\,504
photons. In each panel, there are 4 families of curves from bottom to top with decreasing
gravities, $\log g=4.5,4.0,3.5,3.0$. Results for different chemical compositions are shown
with different types of lines: solar (full), 1/10 solar (dashed), and metal-free (dotted). \label{QQFig}}
\end{figure}

\clearpage

\begin{figure}
\figurenum{16}
\plotone{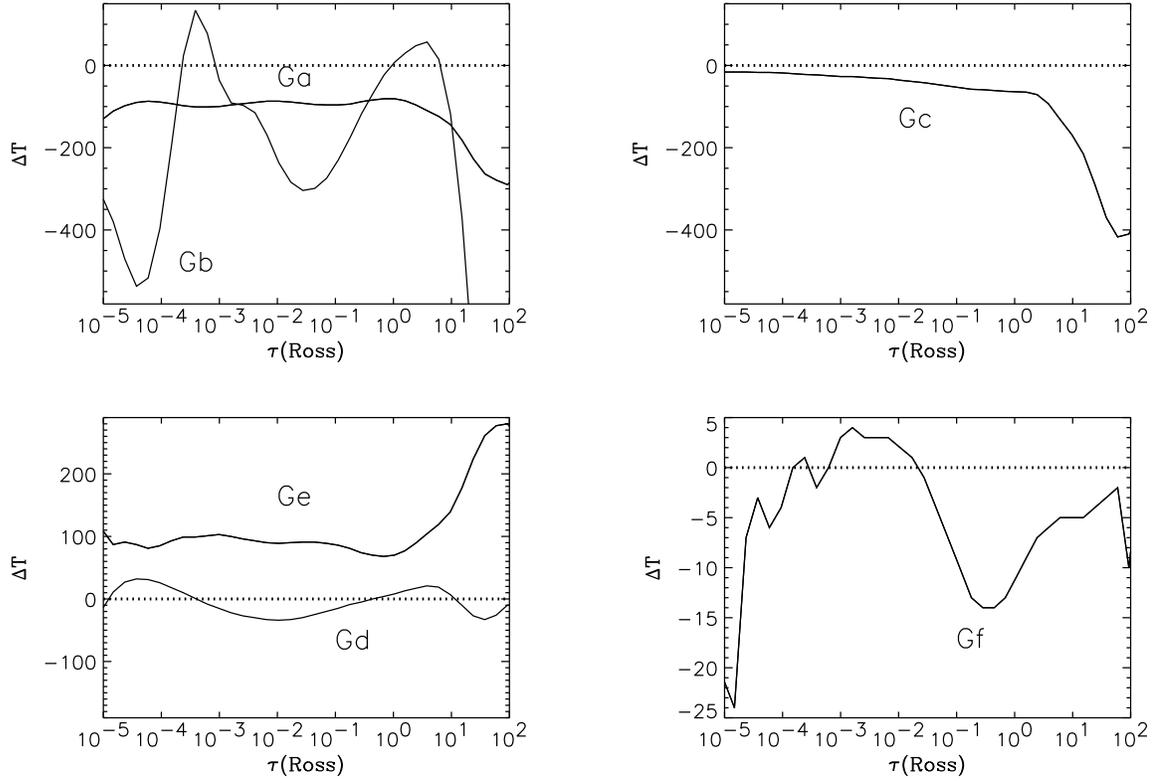}
\caption{Changes in the model temperature structure with respect to the reference
model depending on the treatment of Fe model atoms and line opacity
(see Table~\ref{TestTbl} for model keys). \label{FeTFig}}
\end{figure}

\singlespace
\clearpage

\begin{deluxetable}{lrrrl}
\tabletypesize{\small}
\tablecaption{Atomic data included in the NLTE model atmospheres. \label{IonTbl}}
\tablehead{
\colhead{Ion} & \colhead{(Super)Levels}   & \colhead{Indiv. Levels}   &
\colhead{Lines} & \colhead{References}
}
\startdata
\ion{H}{1}  &   9\phm{Levels}  &  80\phm{Levels}  &     172\phm{a}  &                \\*
\ion{H}{2}  &   1\phm{Levels}  &   1\phm{Levels}  &                 &                \\
&&&&\\
\ion{He}{1} &  24\phm{Levels}  &  72\phm{Levels}  &     784\phm{a}  & \phm{aaa}\citet{NIST}  \\*
\ion{He}{2} &  20\phm{Levels}  &  20\phm{Levels}  &     190\phm{a}  &                \\*
\ion{He}{3} &   1\phm{Levels}  &   1\phm{Levels}  &                 &                \\
&&&&\\
\ion{C}{2}  &  22\phm{Levels}  &  44\phm{Levels}  &     238\phm{a}  & \phm{aaa}\citet{ADOC24} \\*
\ion{C}{3}  &  23\phm{Levels}  &  55\phm{Levels}  &     258\phm{a}  & \phm{aaa}\citet{ADOC14} \\*
\ion{C}{4}  &  25\phm{Levels}  &  55\phm{Levels}  &     330\phm{a}  & \phm{aaa}\citet{ADOC9}  \\*
\ion{C}{5}  &   1\phm{Levels}  &   1\phm{Levels}  &                 &                \\
&&&&\\
\ion{N}{2}  &  26\phm{Levels}  &  71\phm{Levels}  &     373\phm{a}  & \phm{aaa}\citet{ADOC11} \\*
\ion{N}{3}  &  32\phm{Levels}  &  68\phm{Levels}  &     549\phm{a}  & \phm{aaa}\citet{ADOC24} \\*
\ion{N}{4}  &  23\phm{Levels}  &  58\phm{Levels}  &     286\phm{a}  & \phm{aaa}\citet{ADOC14} \\*
\ion{N}{5}  &  16\phm{Levels}  &  55\phm{Levels}  &     330\phm{a}  & \phm{aaa}\citet{ADOC9}  \\*
\ion{N}{6}  &   1\phm{Levels}  &   1\phm{Levels}  &                 &                \\
&&&&\\
\ion{O}{2}  &  29\phm{Levels}  &  62\phm{Levels}  &     346\phm{a}  & \phm{aaa}Burke \& Lennon (unpub.) \\*
\ion{O}{3}  &  29\phm{Levels}  & 116\phm{Levels}  &     846\phm{a}  & \phm{aaa}\citet{ADOC11} \\*
\ion{O}{4}  &  39\phm{Levels}  &  94\phm{Levels}  &     922\phm{a}  & \phm{aaa}\citet{ADOC24} \\*
\ion{O}{5}  &  40\phm{Levels}  &  89\phm{Levels}  &     610\phm{a}  & \phm{aaa}\citet{ADOC14}  \\*
\ion{O}{6}  &  20\phm{Levels}  &  55\phm{Levels}  &     330\phm{a}  & \phm{aaa}\citet{ADOC9}  \\*
\ion{O}{7}  &   1\phm{Levels}  &   1\phm{Levels}  &                 &                \\
&&&&\\
\ion{Ne}{2} &  15\phm{Levels}  &  29\phm{Levels}  &      83\phm{a}  & \phm{aaa}Butler \& Zeippen (unpub.) \\*
\ion{Ne}{3} &  14\phm{Levels}  &  20\phm{Levels}  &      32\phm{a}  & \phm{aaa}Butler \& Zeippen (unpub.) \\*
\ion{Ne}{4} &  12\phm{Levels}  &  18\phm{Levels}  &      38\phm{a}  & \phm{aaa}Burke \& Lennon (unpub.) \\*
\ion{Ne}{5} &   1\phm{Levels}  &   1\phm{Levels}  &                 &                \\
&&&&\\
\ion{Si}{3} &  30\phm{Levels}  & 105\phm{Levels}  &     747\phm{a}  & \phm{aaa}\citet{ADOC19} \\*
\ion{Si}{4} &  23\phm{Levels}  &  53\phm{Levels}  &     306\phm{a}  & \phm{aaa}Taylor (unpub.) \\*
\ion{Si}{5} &   1\phm{Levels}  &   1\phm{Levels}  &                 &                \\
&&&&\\
\ion{P}{4}  &  14\phm{Levels}  &  14\phm{Levels}  &      19\phm{a}  & \phm{aaa}\citet{NIST}  \\*
\ion{P}{5}  &  17\phm{Levels}  &  43\phm{Levels}  &     223\phm{a}  & \phm{aaa}  \\*
\ion{P}{6}  &   1\phm{Levels}  &   1\phm{Levels}  &                 &                \\
&&&&\\
\ion{S}{3}  &  20\phm{Levels}  &  27\phm{Levels}  &      57\phm{a}  & \phm{aaa}\citet{ADOC18} \\*
\ion{S}{4}  &  15\phm{Levels}  &  15\phm{Levels}  &      26\phm{a}  & \phm{aaa}\citet{ADOC23} \\*
\ion{S}{5}  &  12\phm{Levels}  &  12\phm{Levels}  &      11\phm{a}  & \phm{aaa}\citet{ADOC19} \\*
\ion{S}{6}  &  16\phm{Levels}  &  34\phm{Levels}  &     147\phm{a}  & \phm{aaa}Taylor (unpub.) \\*
\ion{S}{7}  &   1\phm{Levels}  &   1\phm{Levels}  &                 &                \\
&&&&\\
\ion{Fe}{3} &  50\phm{Levels}  &12\,660\phm{Levels}  &1\,604\,934\phm{a}  & \phm{aaa}\citet{CD22, NS96} \\*
\ion{Fe}{4} &  43\phm{Levels}  &13\,705\phm{Levels}  &1\,776\,984\phm{a}  & \phm{aaa}\citet{CD22, IRON26} \\*
\ion{Fe}{5} &  42\phm{Levels}  &11\,986\phm{Levels}  &1\,008\,385\phm{a}  & \phm{aaa}\citet{CD22, IRON16} \\*
\ion{Fe}{6} &  32\phm{Levels}  & 4\,740\phm{Levels}  &   475\,750\phm{a}  & \phm{aaa}\citet{CD22} \\*
\ion{Fe}{7} &   1\phm{Levels}  &   1\phm{Levels}  &                 &                \\
&&&&\\
\ion{Ni}{3} &  36\phm{Levels}  &11\,335\phm{Levels}  &1\,309\,729\phm{a}  & \phm{aaa}\citet{CD22} \\*
\ion{Ni}{4} &  38\phm{Levels}  &13\,172\phm{Levels}  &1\,918\,070\phm{a}  & \phm{aaa}\citet{CD22} \\*
\ion{Ni}{5} &  48\phm{Levels}  &13\,184\phm{Levels}  &1\,971\,819\phm{a}  & \phm{aaa}\citet{CD22} \\*
\ion{Ni}{6} &  42\phm{Levels}  &13\,705\phm{Levels}  &2\,259\,798\phm{a}  & \phm{aaa}\citet{CD22} \\*
\ion{Ni}{7} &   1\phm{Levels}  &   1\phm{Levels}  &                 &                \\
\enddata
\end{deluxetable}

\clearpage
\begin{deluxetable}{llcll}
\tablecaption{Key to models chemical compositions.  \label{KeyTbl}}
\tablehead{ \colhead{Key} & \colhead{Composition} & & \colhead{Key} & \colhead{Composition}}
\startdata
 C & $2\times\sun$ &\phm{1/1000}& V & $1/30\times\sun$ \\
 G & $1\times\sun$ &\phm{1/1000}&  W & $1/50\times\sun$\\
 L & $1/2\times\sun$ &\phm{1/1000}&X & $1/100\times\sun$  \\
 S & $1/5\times\sun$ &\phm{1/1000}&  Y & $1/1000\times\sun$ \\
 T & $1/10\times\sun$ &\phm{1/1000}&  Z & $0$
\enddata
\end{deluxetable}

\clearpage
\begin{deluxetable}{llcccccccccc}
\tabletypesize{\small}
\tablecaption{Bolometric corrections as function
   of effective temperature, gravity and metallicity
   (10 metallicities from 2 times solar to metal-free models).  \label{BCTbl}}
\tablehead{
\colhead{\teff\  [K]} & \colhead{$\log g$}   & \multicolumn{10}{c}{BC [mag]} \\
 & & \colhead{2.} & \colhead{1.} & \colhead{0.5} &
\colhead{0.2} & \colhead{0.1} & \colhead{0.033} & \colhead{0.02} & 
\colhead{0.01} & \colhead{0.001} & \colhead{0.}
}
\startdata
 27500 &  3.00 & $-$2.59 &$-$2.62 & $-$2.65 & $-$2.68 & $-$2.71 & $-$2.73 & $-$2.74 & $-$2.75 & $-$2.76 & $-$2.76 \\*
 27500 &  3.25 & $-$2.56 &$-$2.61 & $-$2.65 & $-$2.69 & $-$2.71 & $-$2.73 & $-$2.74 & $-$2.74 & $-$2.75 & $-$2.75 \\*
 27500 &  3.50 & $-$2.57 &$-$2.62 & $-$2.66 & $-$2.70 & $-$2.72 & $-$2.74 & $-$2.75 & $-$2.76 & $-$2.76 & $-$2.76 \\*
 27500 &  3.75 & $-$2.59 &$-$2.64 & $-$2.68 & $-$2.72 & $-$2.74 & $-$2.76 & $-$2.77 & $-$2.77 & $-$2.78 & $-$2.78 \\*
 27500 &  4.00 & $-$2.60 &$-$2.65 & $-$2.69 & $-$2.73 & $-$2.75 & $-$2.77 & $-$2.78 & $-$2.78 & $-$2.79 & $-$2.80 \\*
 27500 &  4.25 & $-$2.62 &$-$2.67 & $-$2.71 & $-$2.74 & $-$2.76 & $-$2.78 & $-$2.79 & $-$2.79 & $-$2.80 & $-$2.81 \\*
 27500 &  4.50 & $-$2.63 &$-$2.68 & $-$2.72 & $-$2.75 & $-$2.77 & $-$2.79 & $-$2.80 & $-$2.80 & $-$2.81 & $-$2.81 \\*
 27500 &  4.75 & $-$2.64 &$-$2.69 & $-$2.72 & $-$2.76 & $-$2.78 & $-$2.80 & $-$2.80 & $-$2.81 & $-$2.82 & $-$2.82 \\[2mm]
 30000 &  3.00 & $-$2.92 &$-$2.92 & $-$2.92 & $-$2.94 & $-$2.95 & $-$2.97 & $-$2.98 & $-$2.99 & $-$3.00 & $-$3.00 \\*
 30000 &  3.25 & $-$2.80 &$-$2.83 & $-$2.86 & $-$2.89 & $-$2.91 & $-$2.93 & $-$2.94 & $-$2.95 & $-$2.96 & $-$2.96 \\*
 30000 &  3.50 & $-$2.78 &$-$2.82 & $-$2.86 & $-$2.90 & $-$2.92 & $-$2.94 & $-$2.94 & $-$2.95 & $-$2.96 & $-$2.96 \\*
 30000 &  3.75 & $-$2.79 &$-$2.84 & $-$2.88 & $-$2.91 & $-$2.93 & $-$2.95 & $-$2.95 & $-$2.96 & $-$2.96 & $-$2.97 \\*
 30000 &  4.00 & $-$2.81 &$-$2.85 & $-$2.89 & $-$2.93 & $-$2.94 & $-$2.96 & $-$2.97 & $-$2.97 & $-$2.98 & $-$2.98 \\*
 30000 &  4.25 & $-$2.82 &$-$2.87 & $-$2.90 & $-$2.94 & $-$2.96 & $-$2.97 & $-$2.98 & $-$2.98 & $-$2.99 & $-$2.99 \\*
 30000 &  4.50 & $-$2.84 &$-$2.88 & $-$2.92 & $-$2.95 & $-$2.97 & $-$2.98 & $-$2.99 & $-$2.99 & $-$3.00 & $-$3.00 \\*
 30000 &  4.75 & $-$2.85 &$-$2.89 & $-$2.93 & $-$2.96 & $-$2.97 & $-$2.99 & $-$3.00 & $-$3.00 & $-$3.01 & $-$3.01 \\[2mm]
 32500 &  3.25 & $-$3.07 &$-$3.08 & $-$3.10 & $-$3.12 & $-$3.14 & $-$3.16 & $-$3.16 & $-$3.17 & $-$3.18 & $-$3.18 \\*
 32500 &  3.50 & $-$3.01 &$-$3.04 & $-$3.06 & $-$3.09 & $-$3.11 & $-$3.13 & $-$3.14 & $-$3.14 & $-$3.15 & $-$3.15 \\*
 32500 &  3.75 & $-$3.00 &$-$3.03 & $-$3.06 & $-$3.09 & $-$3.11 & $-$3.13 & $-$3.13 & $-$3.14 & $-$3.15 & $-$3.15 \\*
 32500 &  4.00 & $-$3.00 &$-$3.04 & $-$3.07 & $-$3.10 & $-$3.12 & $-$3.14 & $-$3.14 & $-$3.14 & $-$3.15 & $-$3.15 \\*
 32500 &  4.25 & $-$3.01 &$-$3.05 & $-$3.08 & $-$3.11 & $-$3.13 & $-$3.14 & $-$3.15 & $-$3.15 & $-$3.16 & $-$3.16 \\*
 32500 &  4.50 & $-$3.03 &$-$3.06 & $-$3.10 & $-$3.12 & $-$3.14 & $-$3.15 & $-$3.15 & $-$3.16 & $-$3.16 & $-$3.16 \\*
 32500 &  4.75 & $-$3.04 &$-$3.08 & $-$3.11 & $-$3.13 & $-$3.15 & $-$3.16 & $-$3.16 & $-$3.16 & $-$3.17 & $-$3.17 \\[2mm]
 35000 &  3.25 & $-$3.35 &$-$3.36 & $-$3.37 & $-$3.38 & $-$3.39 & $-$3.40 & $-$3.41 & $-$3.41 & $-$3.42 & $-$3.42 \\*
 35000 &  3.50 & $-$3.25 &$-$3.27 & $-$3.29 & $-$3.32 & $-$3.33 & $-$3.35 & $-$3.35 & $-$3.35 & $-$3.36 & $-$3.36 \\*
 35000 &  3.75 & $-$3.22 &$-$3.24 & $-$3.27 & $-$3.29 & $-$3.31 & $-$3.33 & $-$3.33 & $-$3.33 & $-$3.34 & $-$3.34 \\*
 35000 &  4.00 & $-$3.21 &$-$3.24 & $-$3.26 & $-$3.29 & $-$3.30 & $-$3.32 & $-$3.32 & $-$3.33 & $-$3.33 & $-$3.33 \\*
 35000 &  4.25 & $-$3.21 &$-$3.24 & $-$3.26 & $-$3.29 & $-$3.30 & $-$3.32 & $-$3.32 & $-$3.32 & $-$3.33 & $-$3.33 \\*
 35000 &  4.50 & $-$3.21 &$-$3.24 & $-$3.27 & $-$3.30 & $-$3.31 & $-$3.32 & $-$3.32 & $-$3.33 & $-$3.33 & $-$3.33 \\*
 35000 &  4.75 & $-$3.22 &$-$3.25 & $-$3.28 & $-$3.30 & $-$3.31 & $-$3.32 & $-$3.32 & $-$3.33 & $-$3.33 & $-$3.33 \\[2mm]
 37500 &  3.50 & $-$3.48 &$-$3.50 & $-$3.52 & $-$3.54 & $-$3.55 & $-$3.57 & $-$3.57 & $-$3.58 & $-$3.58 & $-$3.58 \\*
 37500 &  3.75 & $-$3.44 &$-$3.46 & $-$3.48 & $-$3.51 & $-$3.52 & $-$3.53 & $-$3.54 & $-$3.54 & $-$3.54 & $-$3.55 \\*
 37500 &  4.00 & $-$3.42 &$-$3.45 & $-$3.47 & $-$3.49 & $-$3.50 & $-$3.52 & $-$3.52 & $-$3.52 & $-$3.53 & $-$3.53 \\*
 37500 &  4.25 & $-$3.41 &$-$3.44 & $-$3.46 & $-$3.48 & $-$3.50 & $-$3.51 & $-$3.51 & $-$3.51 & $-$3.52 & $-$3.52 \\*
 37500 &  4.50 & $-$3.41 &$-$3.43 & $-$3.46 & $-$3.48 & $-$3.49 & $-$3.50 & $-$3.51 & $-$3.51 & $-$3.51 & $-$3.51 \\*
 37500 &  4.75 & $-$3.41 &$-$3.44 & $-$3.46 & $-$3.48 & $-$3.49 & $-$3.50 & $-$3.50 & $-$3.51 & $-$3.51 & $-$3.51 \\[2mm]
 40000 &  3.50 & $-$3.72 &$-$3.73 & $-$3.74 & $-$3.76 & $-$3.77 & $-$3.78 & $-$3.79 & $-$3.79 & $-$3.80 & $-$3.80 \\*
 40000 &  3.75 & $-$3.64 &$-$3.67 & $-$3.69 & $-$3.71 & $-$3.73 & $-$3.74 & $-$3.74 & $-$3.75 & $-$3.75 & $-$3.75 \\*
 40000 &  4.00 & $-$3.62 &$-$3.65 & $-$3.67 & $-$3.69 & $-$3.71 & $-$3.72 & $-$3.72 & $-$3.72 & $-$3.73 & $-$3.73 \\*
 40000 &  4.25 & $-$3.61 &$-$3.64 & $-$3.66 & $-$3.68 & $-$3.69 & $-$3.71 & $-$3.71 & $-$3.71 & $-$3.71 & $-$3.71 \\*
 40000 &  4.50 & $-$3.61 &$-$3.63 & $-$3.65 & $-$3.67 & $-$3.69 & $-$3.70 & $-$3.70 & $-$3.70 & $-$3.70 & $-$3.71 \\*
 40000 &  4.75 & $-$3.60 &$-$3.63 & $-$3.65 & $-$3.67 & $-$3.68 & $-$3.69 & $-$3.69 & $-$3.70 & $-$3.70 & $-$3.70 \\[2mm]
 42500 &  3.75 & $-$3.84 &$-$3.86 & $-$3.88 & $-$3.91 & $-$3.92 & $-$3.93 & $-$3.94 & $-$3.94 & $-$3.95 & $-$3.95 \\*
 42500 &  4.00 & $-$3.81 &$-$3.84 & $-$3.86 & $-$3.88 & $-$3.90 & $-$3.91 & $-$3.91 & $-$3.92 & $-$3.92 & $-$3.92 \\*
 42500 &  4.25 & $-$3.80 &$-$3.83 & $-$3.85 & $-$3.87 & $-$3.89 & $-$3.90 & $-$3.90 & $-$3.90 & $-$3.91 & $-$3.91 \\*
 42500 &  4.50 & $-$3.80 &$-$3.82 & $-$3.84 & $-$3.87 & $-$3.88 & $-$3.89 & $-$3.89 & $-$3.89 & $-$3.90 & $-$3.90 \\*
 42500 &  4.75 & $-$3.80 &$-$3.82 & $-$3.84 & $-$3.86 & $-$3.87 & $-$3.88 & $-$3.88 & $-$3.89 & $-$3.89 & $-$3.89 \\[2mm]
 45000 &  3.75 & $-$4.05 &$-$4.06 & $-$4.08 & $-$4.10 & $-$4.11 & $-$4.12 & $-$4.13 & $-$4.13 & $-$4.14 & $-$4.14 \\*
 45000 &  4.00 & $-$3.99 &$-$4.02 & $-$4.04 & $-$4.06 & $-$4.08 & $-$4.09 & $-$4.10 & $-$4.10 & $-$4.11 & $-$4.11 \\*
 45000 &  4.25 & $-$3.98 &$-$4.01 & $-$4.03 & $-$4.05 & $-$4.07 & $-$4.08 & $-$4.08 & $-$4.09 & $-$4.09 & $-$4.09 \\*
 45000 &  4.50 & $-$3.98 &$-$4.00 & $-$4.02 & $-$4.05 & $-$4.06 & $-$4.07 & $-$4.07 & $-$4.08 & $-$4.08 & $-$4.08 \\*
 45000 &  4.75 & $-$3.98 &$-$4.00 & $-$4.02 & $-$4.04 & $-$4.05 & $-$4.07 & $-$4.07 & $-$4.07 & $-$4.08 & $-$4.08 \\[2mm]
 47500 &  3.75 & $-$4.25 &$-$4.26 & $-$4.27 & $-$4.29 & $-$4.30 & $-$4.31 & $-$4.31 & $-$4.32 & $-$4.33 & $-$4.33 \\*
 47500 &  4.00 & $-$4.17 &$-$4.20 & $-$4.22 & $-$4.24 & $-$4.25 & $-$4.26 & $-$4.27 & $-$4.27 & $-$4.28 & $-$4.29 \\*
 47500 &  4.25 & $-$4.15 &$-$4.18 & $-$4.20 & $-$4.22 & $-$4.24 & $-$4.25 & $-$4.25 & $-$4.26 & $-$4.27 & $-$4.27 \\*
 47500 &  4.50 & $-$4.15 &$-$4.17 & $-$4.19 & $-$4.22 & $-$4.23 & $-$4.24 & $-$4.25 & $-$4.25 & $-$4.26 & $-$4.26 \\*
 47500 &  4.75 & $-$4.14 &$-$4.17 & $-$4.19 & $-$4.21 & $-$4.23 & $-$4.24 & $-$4.24 & $-$4.25 & $-$4.25 & $-$4.25 \\[2mm]
 50000 &  4.00 & $-$4.35 &$-$4.37 & $-$4.39 & $-$4.41 & $-$4.42 & $-$4.43 & $-$4.44 & $-$4.44 & $-$4.45 & $-$4.45 \\*
 50000 &  4.25 & $-$4.32 &$-$4.34 & $-$4.36 & $-$4.39 & $-$4.40 & $-$4.41 & $-$4.42 & $-$4.42 & $-$4.43 & $-$4.43 \\*
 50000 &  4.50 & $-$4.31 &$-$4.33 & $-$4.36 & $-$4.38 & $-$4.39 & $-$4.41 & $-$4.41 & $-$4.41 & $-$4.42 & $-$4.42 \\*
 50000 &  4.75 & $-$4.30 &$-$4.33 & $-$4.35 & $-$4.38 & $-$4.39 & $-$4.40 & $-$4.41 & $-$4.41 & $-$4.42 & $-$4.42 \\[2mm]
 52500 &  4.00 & $-$4.52 &$-$4.54 & $-$4.55 & $-$4.57 & $-$4.58 & $-$4.60 & $-$4.60 & $-$4.60 & $-$4.61 & $-$4.61 \\*
 52500 &  4.25 & $-$4.48 &$-$4.50 & $-$4.52 & $-$4.54 & $-$4.56 & $-$4.57 & $-$4.58 & $-$4.58 & $-$4.59 & $-$4.59 \\*
 52500 &  4.50 & $-$4.46 &$-$4.49 & $-$4.51 & $-$4.53 & $-$4.55 & $-$4.56 & $-$4.57 & $-$4.57 & $-$4.58 & $-$4.58 \\*
 52500 &  4.75 & $-$4.46 &$-$4.49 & $-$4.51 & $-$4.53 & $-$4.54 & $-$4.56 & $-$4.56 & $-$4.57 & $-$4.57 & $-$4.58 \\[2mm]
 55000 &  4.00 & $-$4.69 &$-$4.71 & $-$4.72 & $-$4.74 & $-$4.75 & $-$4.76 & $-$4.76 & $-$4.76 & $-$4.77 & $-$4.77 \\*
 55000 &  4.25 & $-$4.63 &$-$4.66 & $-$4.68 & $-$4.70 & $-$4.71 & $-$4.72 & $-$4.73 & $-$4.73 & $-$4.74 & $-$4.74 \\*
 55000 &  4.50 & $-$4.62 &$-$4.64 & $-$4.66 & $-$4.69 & $-$4.70 & $-$4.71 & $-$4.72 & $-$4.72 & $-$4.73 & $-$4.73 \\*
 55000 &  4.75 & $-$4.61 &$-$4.64 & $-$4.66 & $-$4.68 & $-$4.69 & $-$4.71 & $-$4.71 & $-$4.72 & $-$4.72 & $-$4.73
\enddata
\end{deluxetable}

\clearpage
\begin{deluxetable}{llcccccccccc}
\tabletypesize{\small}
\tablecaption{Ionizing fluxes in the \ion{H}{1} Lyman continuum as function
   of effective temperature, gravity and metallicity
   (10 metallicities from 2 times solar to metal-free models).  \label{Q0Tbl}}
\tablehead{
\colhead{\teff\  [K]} & \colhead{$\log g$}   & \multicolumn{10}{c}{
$q_0 = \log N_{\rm LyC}$ [s$^{-1}$ cm$^{-2}$]} \\
 & & \colhead{2.} & \colhead{1.} & \colhead{0.5} &
\colhead{0.2} & \colhead{0.1} & \colhead{0.033} & \colhead{0.02} & 
\colhead{0.01} & \colhead{0.001} & \colhead{0.}
}
\startdata
 27500 &  3.00 &  23.26  & 23.21 &  23.18 &  23.15 &  23.14 &  23.13 &  23.13 &  23.13 &  23.14 &  23.15 \\*
 27500 &  3.25 &  23.00  & 22.93 &  22.87 &  22.81 &  22.78 &  22.76 &  22.75 &  22.74 &  22.74 &  22.75 \\*
 27500 &  3.50 &  22.79  & 22.70 &  22.62 &  22.53 &  22.48 &  22.41 &  22.39 &  22.37 &  22.33 &  22.33 \\*
 27500 &  3.75 &  22.62  & 22.52 &  22.43 &  22.33 &  22.26 &  22.17 &  22.14 &  22.10 &  22.03 &  22.01 \\*
 27500 &  4.00 &  22.49  & 22.39 &  22.30 &  22.19 &  22.12 &  22.01 &  21.97 &  21.93 &  21.84 &  21.82 \\*
 27500 &  4.25 &  22.39  & 22.28 &  22.19 &  22.08 &  22.01 &  21.90 &  21.86 &  21.81 &  21.73 &  21.70 \\*
 27500 &  4.50 &  22.30  & 22.20 &  22.11 &  22.00 &  21.92 &  21.82 &  21.78 &  21.74 &  21.66 &  21.64 \\*
 27500 &  4.75 &  22.23  & 22.13 &  22.03 &  21.92 &  21.85 &  21.76 &  21.72 &  21.68 &  21.61 &  21.59 \\[2mm]
 30000 &  3.00 &  23.68  & 23.66 &  23.65 &  23.64 &  23.63 &  23.63 &  23.63 &  23.63 &  23.63 &  23.63 \\*
 30000 &  3.25 &  23.51  & 23.47 &  23.44 &  23.42 &  23.41 &  23.40 &  23.40 &  23.40 &  23.41 &  23.41 \\*
 30000 &  3.50 &  23.36  & 23.30 &  23.26 &  23.21 &  23.19 &  23.18 &  23.17 &  23.17 &  23.18 &  23.18 \\*
 30000 &  3.75 &  23.22  & 23.15 &  23.08 &  23.02 &  22.99 &  22.96 &  22.95 &  22.95 &  22.94 &  22.95 \\*
 30000 &  4.00 &  23.10  & 23.01 &  22.94 &  22.86 &  22.81 &  22.76 &  22.75 &  22.74 &  22.73 &  22.73 \\*
 30000 &  4.25 &  22.99  & 22.89 &  22.81 &  22.72 &  22.67 &  22.61 &  22.59 &  22.58 &  22.56 &  22.55 \\*
 30000 &  4.50 &  22.90  & 22.80 &  22.71 &  22.62 &  22.56 &  22.50 &  22.48 &  22.46 &  22.43 &  22.43 \\*
 30000 &  4.75 &  22.82  & 22.72 &  22.63 &  22.54 &  22.49 &  22.42 &  22.40 &  22.38 &  22.34 &  22.34 \\[2mm]
 32500 &  3.25 &  23.84  & 23.82 &  23.81 &  23.79 &  23.79 &  23.78 &  23.78 &  23.78 &  23.78 &  23.78 \\*
 32500 &  3.50 &  23.73  & 23.70 &  23.68 &  23.66 &  23.65 &  23.64 &  23.64 &  23.64 &  23.64 &  23.64 \\*
 32500 &  3.75 &  23.64  & 23.60 &  23.56 &  23.53 &  23.52 &  23.51 &  23.51 &  23.51 &  23.51 &  23.51 \\*
 32500 &  4.00 &  23.55  & 23.50 &  23.45 &  23.41 &  23.39 &  23.38 &  23.38 &  23.38 &  23.38 &  23.38 \\*
 32500 &  4.25 &  23.47  & 23.41 &  23.36 &  23.31 &  23.28 &  23.26 &  23.26 &  23.26 &  23.26 &  23.26 \\*
 32500 &  4.50 &  23.40  & 23.33 &  23.27 &  23.21 &  23.18 &  23.16 &  23.15 &  23.15 &  23.15 &  23.15 \\*
 32500 &  4.75 &  23.34  & 23.26 &  23.20 &  23.13 &  23.10 &  23.07 &  23.07 &  23.06 &  23.06 &  23.06 \\[2mm]
 35000 &  3.25 &  24.09  & 24.08 &  24.07 &  24.06 &  24.06 &  24.05 &  24.05 &  24.05 &  24.05 &  24.05 \\*
 35000 &  3.50 &  23.99  & 23.98 &  23.97 &  23.95 &  23.95 &  23.94 &  23.94 &  23.94 &  23.94 &  23.94 \\*
 35000 &  3.75 &  23.92  & 23.90 &  23.88 &  23.87 &  23.86 &  23.85 &  23.85 &  23.85 &  23.85 &  23.85 \\*
 35000 &  4.00 &  23.87  & 23.84 &  23.81 &  23.79 &  23.78 &  23.77 &  23.77 &  23.77 &  23.77 &  23.77 \\*
 35000 &  4.25 &  23.81  & 23.78 &  23.75 &  23.72 &  23.71 &  23.70 &  23.70 &  23.70 &  23.70 &  23.70 \\*
 35000 &  4.50 &  23.77  & 23.72 &  23.69 &  23.66 &  23.64 &  23.64 &  23.64 &  23.64 &  23.64 &  23.64 \\*
 35000 &  4.75 &  23.72  & 23.67 &  23.64 &  23.60 &  23.59 &  23.58 &  23.58 &  23.58 &  23.58 &  23.58 \\[2mm]
 37500 &  3.50 &  24.21  & 24.20 &  24.19 &  24.18 &  24.17 &  24.17 &  24.17 &  24.16 &  24.16 &  24.16 \\*
 37500 &  3.75 &  24.14  & 24.13 &  24.12 &  24.11 &  24.10 &  24.10 &  24.09 &  24.09 &  24.09 &  24.09 \\*
 37500 &  4.00 &  24.10  & 24.08 &  24.07 &  24.05 &  24.04 &  24.04 &  24.04 &  24.04 &  24.04 &  24.04 \\*
 37500 &  4.25 &  24.06  & 24.04 &  24.02 &  24.00 &  24.00 &  23.99 &  23.99 &  23.99 &  23.99 &  23.99 \\*
 37500 &  4.50 &  24.03  & 24.00 &  23.98 &  23.96 &  23.96 &  23.95 &  23.95 &  23.95 &  23.95 &  23.95 \\*
 37500 &  4.75 &  24.00  & 23.97 &  23.95 &  23.93 &  23.92 &  23.92 &  23.91 &  23.91 &  23.92 &  23.92 \\[2mm]
 40000 &  3.50 &  24.39  & 24.38 &  24.37 &  24.36 &  24.36 &  24.35 &  24.35 &  24.35 &  24.34 &  24.34 \\*
 40000 &  3.75 &  24.33  & 24.32 &  24.31 &  24.30 &  24.30 &  24.29 &  24.29 &  24.28 &  24.28 &  24.28 \\*
 40000 &  4.00 &  24.29  & 24.28 &  24.27 &  24.26 &  24.25 &  24.24 &  24.24 &  24.24 &  24.24 &  24.24 \\*
 40000 &  4.25 &  24.26  & 24.24 &  24.23 &  24.22 &  24.21 &  24.20 &  24.20 &  24.20 &  24.20 &  24.20 \\*
 40000 &  4.50 &  24.23  & 24.21 &  24.20 &  24.19 &  24.18 &  24.18 &  24.17 &  24.17 &  24.17 &  24.17 \\*
 40000 &  4.75 &  24.21  & 24.19 &  24.18 &  24.16 &  24.16 &  24.15 &  24.15 &  24.15 &  24.15 &  24.15 \\[2mm]
 42500 &  3.75 &  24.49  & 24.48 &  24.47 &  24.46 &  24.46 &  24.45 &  24.45 &  24.45 &  24.44 &  24.44 \\*
 42500 &  4.00 &  24.45  & 24.44 &  24.43 &  24.42 &  24.42 &  24.41 &  24.41 &  24.40 &  24.40 &  24.40 \\*
 42500 &  4.25 &  24.43  & 24.41 &  24.40 &  24.39 &  24.38 &  24.38 &  24.38 &  24.37 &  24.37 &  24.37 \\*
 42500 &  4.50 &  24.40  & 24.39 &  24.38 &  24.37 &  24.36 &  24.35 &  24.35 &  24.35 &  24.35 &  24.35 \\*
 42500 &  4.75 &  24.39  & 24.37 &  24.36 &  24.35 &  24.34 &  24.34 &  24.33 &  24.33 &  24.33 &  24.33 \\[2mm]
 45000 &  3.75 &  24.62  & 24.62 &  24.61 &  24.60 &  24.60 &  24.59 &  24.59 &  24.59 &  24.58 &  24.58 \\*
 45000 &  4.00 &  24.59  & 24.58 &  24.58 &  24.57 &  24.56 &  24.55 &  24.55 &  24.55 &  24.55 &  24.54 \\*
 45000 &  4.25 &  24.57  & 24.56 &  24.55 &  24.54 &  24.53 &  24.53 &  24.52 &  24.52 &  24.52 &  24.52 \\*
 45000 &  4.50 &  24.55  & 24.54 &  24.53 &  24.52 &  24.51 &  24.50 &  24.50 &  24.50 &  24.50 &  24.50 \\*
 45000 &  4.75 &  24.54  & 24.52 &  24.51 &  24.50 &  24.50 &  24.49 &  24.49 &  24.48 &  24.48 &  24.48 \\[2mm]
 47500 &  3.75 &  24.74  & 24.74 &  24.73 &  24.72 &  24.72 &  24.71 &  24.71 &  24.71 &  24.71 &  24.71 \\*
 47500 &  4.00 &  24.72  & 24.71 &  24.70 &  24.69 &  24.69 &  24.68 &  24.68 &  24.68 &  24.67 &  24.67 \\*
 47500 &  4.25 &  24.70  & 24.69 &  24.68 &  24.67 &  24.66 &  24.65 &  24.65 &  24.65 &  24.65 &  24.65 \\*
 47500 &  4.50 &  24.68  & 24.67 &  24.66 &  24.65 &  24.64 &  24.64 &  24.63 &  24.63 &  24.63 &  24.63 \\*
 47500 &  4.75 &  24.67  & 24.66 &  24.65 &  24.64 &  24.63 &  24.62 &  24.62 &  24.62 &  24.61 &  24.61 \\[2mm]
 50000 &  4.00 &  24.82  & 24.82 &  24.81 &  24.80 &  24.80 &  24.79 &  24.79 &  24.79 &  24.79 &  24.78 \\*
 50000 &  4.25 &  24.81  & 24.80 &  24.79 &  24.78 &  24.78 &  24.77 &  24.77 &  24.77 &  24.76 &  24.76 \\*
 50000 &  4.50 &  24.80  & 24.79 &  24.78 &  24.77 &  24.76 &  24.75 &  24.75 &  24.75 &  24.74 &  24.74 \\*
 50000 &  4.75 &  24.79  & 24.78 &  24.76 &  24.75 &  24.75 &  24.74 &  24.74 &  24.74 &  24.73 &  24.73 \\[2mm]
 52500 &  4.00 &  24.92  & 24.92 &  24.91 &  24.90 &  24.90 &  24.90 &  24.89 &  24.89 &  24.89 &  24.89 \\*
 52500 &  4.25 &  24.91  & 24.90 &  24.89 &  24.89 &  24.88 &  24.87 &  24.87 &  24.87 &  24.87 &  24.87 \\*
 52500 &  4.50 &  24.90  & 24.89 &  24.88 &  24.87 &  24.87 &  24.86 &  24.86 &  24.85 &  24.85 &  24.85 \\*
 52500 &  4.75 &  24.89  & 24.88 &  24.87 &  24.86 &  24.85 &  24.85 &  24.84 &  24.84 &  24.84 &  24.84 \\[2mm]
 55000 &  4.00 &  25.01  & 25.01 &  25.00 &  25.00 &  24.99 &  24.99 &  24.99 &  24.99 &  24.98 &  24.98 \\*
 55000 &  4.25 &  25.00  & 24.99 &  24.99 &  24.98 &  24.98 &  24.97 &  24.97 &  24.97 &  24.96 &  24.96 \\*
 55000 &  4.50 &  24.99  & 24.98 &  24.97 &  24.97 &  24.96 &  24.96 &  24.95 &  24.95 &  24.95 &  24.95 \\*
 55000 &  4.75 &  24.98  & 24.97 &  24.97 &  24.96 &  24.95 &  24.94 &  24.94 &  24.94 &  24.94 &  24.94
\enddata
\end{deluxetable}

\clearpage
\begin{deluxetable}{llcccccccccc}
\tabletypesize{\small}
\tablecaption{Ionizing fluxes in the \ion{He}{1} continuum as function
   of effective temperature, gravity and metallicity
   (10 metallicities from 2 times solar to metal-free models).  \label{Q1Tbl}}
\tablehead{
\colhead{\teff\  [K]} & \colhead{$\log g$}   & \multicolumn{10}{c}{
$q_1 = \log N_{504}$ [s$^{-1}$ cm$^{-2}$]} \\
 & & \colhead{2.} & \colhead{1.} & \colhead{0.5} &
\colhead{0.2} & \colhead{0.1} & \colhead{0.033} & \colhead{0.02} & 
\colhead{0.01} & \colhead{0.001} & \colhead{0.}
}
\startdata
 27500 &  3.00 &  20.85  & 20.80 &  20.74 &  20.69 &  20.65 &  20.61 &  20.60 &  20.58 &  20.55 &  20.53 \\*
 27500 &  3.25 &  20.45  & 20.35 &  20.25 &  20.10 &  19.99 &  19.82 &  19.76 &  19.71 &  19.62 &  19.60 \\*
 27500 &  3.50 &  20.17  & 20.03 &  19.89 &  19.68 &  19.52 &  19.30 &  19.21 &  19.13 &  19.00 &  18.98 \\*
 27500 &  3.75 &  19.95  & 19.80 &  19.64 &  19.40 &  19.22 &  18.95 &  18.86 &  18.76 &  18.60 &  18.57 \\*
 27500 &  4.00 &  19.79  & 19.63 &  19.45 &  19.20 &  19.01 &  18.73 &  18.63 &  18.53 &  18.37 &  18.34 \\*
 27500 &  4.25 &  19.66  & 19.49 &  19.31 &  19.05 &  18.85 &  18.58 &  18.48 &  18.39 &  18.24 &  18.21 \\*
 27500 &  4.50 &  19.55  & 19.37 &  19.18 &  18.92 &  18.72 &  18.46 &  18.37 &  18.29 &  18.16 &  18.13 \\*
 27500 &  4.75 &  19.45  & 19.26 &  19.07 &  18.80 &  18.61 &  18.37 &  18.29 &  18.22 &  18.11 &  18.08 \\[2mm]
 30000 &  3.00 &  22.29  & 22.31 &  22.33 &  22.35 &  22.38 &  22.41 &  22.43 &  22.45 &  22.47 &  22.48 \\*
 30000 &  3.25 &  21.55  & 21.51 &  21.50 &  21.52 &  21.56 &  21.61 &  21.63 &  21.66 &  21.70 &  21.70 \\*
 30000 &  3.50 &  21.12  & 21.06 &  21.00 &  20.94 &  20.89 &  20.84 &  20.82 &  20.81 &  20.78 &  20.77 \\*
 30000 &  3.75 &  20.90  & 20.81 &  20.71 &  20.58 &  20.49 &  20.36 &  20.32 &  20.28 &  20.22 &  20.21 \\*
 30000 &  4.00 &  20.73  & 20.61 &  20.48 &  20.31 &  20.18 &  20.02 &  19.97 &  19.92 &  19.85 &  19.83 \\*
 30000 &  4.25 &  20.59  & 20.45 &  20.30 &  20.10 &  19.96 &  19.77 &  19.71 &  19.66 &  19.58 &  19.56 \\*
 30000 &  4.50 &  20.46  & 20.31 &  20.15 &  19.94 &  19.79 &  19.60 &  19.54 &  19.47 &  19.39 &  19.37 \\*
 30000 &  4.75 &  20.36  & 20.20 &  20.04 &  19.82 &  19.67 &  19.47 &  19.41 &  19.35 &  19.27 &  19.25 \\[2mm]
 32500 &  3.25 &  22.61  & 22.64 &  22.67 &  22.70 &  22.73 &  22.77 &  22.78 &  22.80 &  22.82 &  22.83 \\*
 32500 &  3.50 &  22.24  & 22.25 &  22.27 &  22.30 &  22.34 &  22.39 &  22.41 &  22.43 &  22.47 &  22.48 \\*
 32500 &  3.75 &  21.87  & 21.85 &  21.85 &  21.87 &  21.90 &  21.95 &  21.97 &  21.99 &  22.04 &  22.05 \\*
 32500 &  4.00 &  21.61  & 21.56 &  21.53 &  21.51 &  21.51 &  21.51 &  21.52 &  21.53 &  21.56 &  21.56 \\*
 32500 &  4.25 &  21.44  & 21.38 &  21.32 &  21.25 &  21.21 &  21.17 &  21.16 &  21.15 &  21.15 &  21.14 \\*
 32500 &  4.50 &  21.32  & 21.24 &  21.15 &  21.05 &  20.99 &  20.92 &  20.90 &  20.88 &  20.86 &  20.85 \\*
 32500 &  4.75 &  21.22  & 21.12 &  21.02 &  20.90 &  20.82 &  20.73 &  20.71 &  20.69 &  20.66 &  20.65 \\[2mm]
 35000 &  3.25 &  23.12  & 23.15 &  23.17 &  23.21 &  23.23 &  23.26 &  23.27 &  23.28 &  23.30 &  23.31 \\*
 35000 &  3.50 &  22.92  & 22.96 &  22.99 &  23.03 &  23.05 &  23.09 &  23.10 &  23.12 &  23.14 &  23.15 \\*
 35000 &  3.75 &  22.75  & 22.77 &  22.80 &  22.83 &  22.86 &  22.91 &  22.93 &  22.95 &  22.98 &  22.98 \\*
 35000 &  4.00 &  22.56  & 22.58 &  22.60 &  22.63 &  22.66 &  22.71 &  22.74 &  22.76 &  22.80 &  22.80 \\*
 35000 &  4.25 &  22.38  & 22.38 &  22.40 &  22.43 &  22.46 &  22.51 &  22.54 &  22.56 &  22.60 &  22.61 \\*
 35000 &  4.50 &  22.21  & 22.21 &  22.22 &  22.24 &  22.27 &  22.32 &  22.34 &  22.37 &  22.41 &  22.41 \\*
 35000 &  4.75 &  22.08  & 22.07 &  22.06 &  22.08 &  22.10 &  22.13 &  22.15 &  22.18 &  22.22 &  22.22 \\[2mm]
 37500 &  3.50 &  23.30  & 23.34 &  23.37 &  23.41 &  23.43 &  23.46 &  23.47 &  23.48 &  23.50 &  23.51 \\*
 37500 &  3.75 &  23.21  & 23.24 &  23.27 &  23.31 &  23.34 &  23.37 &  23.39 &  23.40 &  23.42 &  23.42 \\*
 37500 &  4.00 &  23.12  & 23.15 &  23.18 &  23.21 &  23.24 &  23.28 &  23.30 &  23.31 &  23.34 &  23.34 \\*
 37500 &  4.25 &  23.03  & 23.06 &  23.08 &  23.12 &  23.15 &  23.19 &  23.21 &  23.23 &  23.25 &  23.26 \\*
 37500 &  4.50 &  22.94  & 22.97 &  22.99 &  23.03 &  23.05 &  23.10 &  23.12 &  23.14 &  23.17 &  23.17 \\*
 37500 &  4.75 &  22.85  & 22.87 &  22.90 &  22.94 &  22.97 &  23.02 &  23.04 &  23.06 &  23.08 &  23.09 \\[2mm]
 40000 &  3.50 &  23.60  & 23.62 &  23.64 &  23.67 &  23.69 &  23.72 &  23.72 &  23.73 &  23.75 &  23.75 \\*
 40000 &  3.75 &  23.51  & 23.54 &  23.57 &  23.61 &  23.63 &  23.66 &  23.67 &  23.68 &  23.70 &  23.70 \\*
 40000 &  4.00 &  23.46  & 23.49 &  23.52 &  23.56 &  23.58 &  23.62 &  23.63 &  23.64 &  23.66 &  23.66 \\*
 40000 &  4.25 &  23.41  & 23.44 &  23.47 &  23.51 &  23.54 &  23.57 &  23.59 &  23.60 &  23.61 &  23.62 \\*
 40000 &  4.50 &  23.37  & 23.40 &  23.42 &  23.46 &  23.49 &  23.53 &  23.54 &  23.55 &  23.57 &  23.58 \\*
 40000 &  4.75 &  23.32  & 23.35 &  23.38 &  23.42 &  23.44 &  23.49 &  23.50 &  23.51 &  23.53 &  23.53 \\[2mm]
 42500 &  3.75 &  23.75  & 23.77 &  23.80 &  23.83 &  23.85 &  23.87 &  23.88 &  23.89 &  23.91 &  23.91 \\*
 42500 &  4.00 &  23.70  & 23.73 &  23.76 &  23.80 &  23.82 &  23.85 &  23.86 &  23.87 &  23.88 &  23.89 \\*
 42500 &  4.25 &  23.67  & 23.71 &  23.74 &  23.77 &  23.80 &  23.83 &  23.84 &  23.85 &  23.86 &  23.87 \\*
 42500 &  4.50 &  23.65  & 23.68 &  23.71 &  23.75 &  23.78 &  23.81 &  23.82 &  23.83 &  23.84 &  23.84 \\*
 42500 &  4.75 &  23.63  & 23.66 &  23.69 &  23.73 &  23.75 &  23.79 &  23.80 &  23.81 &  23.82 &  23.82 \\[2mm]
 45000 &  3.75 &  23.96  & 23.98 &  24.00 &  24.03 &  24.04 &  24.06 &  24.06 &  24.07 &  24.08 &  24.08 \\*
 45000 &  4.00 &  23.91  & 23.94 &  23.96 &  23.99 &  24.01 &  24.03 &  24.04 &  24.05 &  24.06 &  24.07 \\*
 45000 &  4.25 &  23.88  & 23.91 &  23.94 &  23.98 &  24.00 &  24.02 &  24.03 &  24.04 &  24.05 &  24.05 \\*
 45000 &  4.50 &  23.87  & 23.90 &  23.93 &  23.97 &  23.99 &  24.01 &  24.02 &  24.03 &  24.04 &  24.04 \\*
 45000 &  4.75 &  23.86  & 23.89 &  23.92 &  23.95 &  23.98 &  24.00 &  24.01 &  24.02 &  24.03 &  24.03 \\[2mm]
 47500 &  3.75 &  24.17  & 24.18 &  24.19 &  24.21 &  24.21 &  24.22 &  24.23 &  24.23 &  24.24 &  24.24 \\*
 47500 &  4.00 &  24.09  & 24.12 &  24.14 &  24.16 &  24.18 &  24.19 &  24.20 &  24.21 &  24.22 &  24.22 \\*
 47500 &  4.25 &  24.07  & 24.09 &  24.12 &  24.15 &  24.16 &  24.18 &  24.19 &  24.20 &  24.21 &  24.21 \\*
 47500 &  4.50 &  24.05  & 24.08 &  24.11 &  24.14 &  24.16 &  24.18 &  24.19 &  24.19 &  24.20 &  24.21 \\*
 47500 &  4.75 &  24.04  & 24.07 &  24.10 &  24.14 &  24.16 &  24.18 &  24.18 &  24.19 &  24.20 &  24.20 \\[2mm]
 50000 &  4.00 &  24.26  & 24.28 &  24.30 &  24.32 &  24.33 &  24.34 &  24.34 &  24.35 &  24.35 &  24.36 \\*
 50000 &  4.25 &  24.23  & 24.26 &  24.28 &  24.30 &  24.31 &  24.33 &  24.33 &  24.34 &  24.35 &  24.35 \\*
 50000 &  4.50 &  24.22  & 24.24 &  24.27 &  24.29 &  24.31 &  24.32 &  24.33 &  24.34 &  24.35 &  24.35 \\*
 50000 &  4.75 &  24.21  & 24.24 &  24.26 &  24.29 &  24.31 &  24.32 &  24.33 &  24.34 &  24.35 &  24.35 \\[2mm]
 52500 &  4.00 &  24.42  & 24.43 &  24.44 &  24.46 &  24.46 &  24.47 &  24.47 &  24.47 &  24.48 &  24.48 \\*
 52500 &  4.25 &  24.38  & 24.40 &  24.42 &  24.44 &  24.45 &  24.46 &  24.46 &  24.47 &  24.47 &  24.47 \\*
 52500 &  4.50 &  24.37  & 24.39 &  24.41 &  24.43 &  24.44 &  24.45 &  24.46 &  24.46 &  24.47 &  24.47 \\*
 52500 &  4.75 &  24.36  & 24.38 &  24.40 &  24.43 &  24.44 &  24.46 &  24.46 &  24.47 &  24.47 &  24.48 \\[2mm]
 55000 &  4.00 &  24.56  & 24.57 &  24.58 &  24.58 &  24.59 &  24.59 &  24.59 &  24.59 &  24.60 &  24.60 \\*
 55000 &  4.25 &  24.52  & 24.54 &  24.55 &  24.56 &  24.57 &  24.58 &  24.58 &  24.58 &  24.59 &  24.59 \\*
 55000 &  4.50 &  24.50  & 24.52 &  24.54 &  24.55 &  24.56 &  24.57 &  24.58 &  24.58 &  24.59 &  24.59 \\*
 55000 &  4.75 &  24.49  & 24.52 &  24.53 &  24.55 &  24.56 &  24.57 &  24.58 &  24.58 &  24.59 &  24.59
\enddata
\end{deluxetable}

\clearpage
\begin{deluxetable}{llrrr}
\tablecaption{Assumptions and characteristics for the test case models.  \label{TestTbl}}
\tablehead{ \colhead{Model} & \colhead{Characteristics} & \colhead{Levels} &
   \colhead{Freqs} & \colhead{Fe Lines}}
\startdata
 G35000g400v10 & Reference model (OS, 0.75) & 907 & 184\,136 & 1\,176\,853 \\
 Ga35       & OS, 30 Doppler widths & 907 & 65\,398 & 1\,176\,821 \\
 Gb35       & ODF & 907 & 40\,759 & 7\,459\,416 \\
 Gc35       & Fewer selected Fe/Ni lines & 907 & 184\,136 & 88\,009 \\
 Gd35       & More Fe {\sc iv-v} superlevels & 965 & 65\,493 & 1\,221\,466 \\
 Ge35       & Less Fe {\sc iv-v} superlevels & 874 & 65\,366 & 952\,051 \\
 Gf35       & No levels above ioniz. limit & 804 & 184\,226 & 1\,168\,764 \\
\enddata
\end{deluxetable}

\end{document}